\documentclass{aa}  

\usepackage{graphicx}
\usepackage{comment}
\usepackage{soul} 

\usepackage{txfonts}
\usepackage[colorlinks=true,
    linkcolor=blue,
    filecolor=magenta,      
    urlcolor=cyan,
    citecolor=blue]{hyperref}

\usepackage{xcolor}
\makeatletter
\renewcommand*\aa@pageof{, page \thepage{} of \pageref*{LastPage}}
\makeatother

\newcommand{\OfourALVKduty}{$15.0^{+15.4}_{-8.8}$}

\newcommand{\OfourBLVKduty}{$16.1^{+16.6}_{-9.4}$}
\newcommand{\OfourCLVKduty}{$19.0^{+19.5}_{-11.1}$}
\newcommand{\OfourDLVKduty}{$18.0^{+18.5}_{-10.5}$}
\newcommand{\OfiveALVKIduty}{$140^{+143}_{-81}$}

\newcommand{\OfiveBLVKduty}{$136^{+140}_{-80}$}
\newcommand{\OfiveCLVKduty}{$144^{+148}_{-84}$}
\newcommand{\OfiveDLVKduty}{$144^{+147}_{-84}$}

\newcommand{\OfourAkng}{$0.16^{+0.17}_{-0.10}$}
\newcommand{\OfourAknz}{$0.16^{+0.17}_{-0.10}$}
\newcommand{\OfourAknJ}{$0.05^{+0.05}_{-0.03}$}
\newcommand{\OfourBkng}{$0.53^{+0.55}_{-0.31}$}
\newcommand{\OfourBknz}{$0.54^{+0.55}_{-0.31}$}
\newcommand{\OfourBknJ}{$0.29^{+0.30}_{-0.17}$}
\newcommand{\OfourCkng}{$0.54^{+0.56}_{-0.32}$}
\newcommand{\OfourCknz}{$0.56^{+0.57}_{-0.33}$}
\newcommand{\OfourCknJ}{$0.30^{+0.31}_{-0.18}$}
\newcommand{\OfourDkng}{$1.55^{+1.60}_{-0.91}$}
\newcommand{\OfourDknz}{$1.57^{+1.61}_{-0.92}$}
\newcommand{\OfourDknJ}{$0.97^{+0.99}_{-0.56}$}
\newcommand{\OfiveAkng}{$2.1^{+2.1}_{-1.2}$}
\newcommand{\OfiveAknz}{$1.5^{+1.6}_{-0.9}$}
\newcommand{\OfiveAknJ}{$0.08^{+0.08}_{-0.05}$}
\newcommand{\OfiveBkng}{$7.0^{+7.2}_{-4.1}$}
\newcommand{\OfiveBknz}{$5.6^{+5.8}_{-3.3}$}
\newcommand{\OfiveBknJ}{$0.48^{+0.49}_{-0.28}$}
\newcommand{\OfiveCkng}{$7.9^{+8.1}_{-4.6}$}
\newcommand{\OfiveCknz}{$6.8^{+6.9}_{-3.9}$}
\newcommand{\OfiveCknJ}{$0.53^{+0.54}_{-0.31}$}
\newcommand{\OfiveDkng}{$15.9^{+16.4}_{-9.3}$}
\newcommand{\OfiveDknz}{$13.4^{+13.8}_{-7.8}$}
\newcommand{\OfiveDknJ}{$1.6^{+1.7}_{-0.9}$}

\newcommand{\OfourAafterradio}{$0.015^{+0.015}_{-0.009}$}
\newcommand{\OfourAafteroptic}{$0.002^{+0.002}_{-0.001}$}
\newcommand{\OfourAafterx}{$0.02^{+0.02}_{-0.01}$}
\newcommand{\OfourAafterradiodeg}{$0.011^{+0.011}_{-0.006}$}
\newcommand{\OfourAafteropticdeg}{$0.001^{+0.001}_{-0.001}$}
\newcommand{\OfourAafterxdeg}{$0.05^{+0.05}_{-0.03}$}
\newcommand{\OfourBafterradio}{$0.06^{+0.06}_{-0.03}$}
\newcommand{\OfourBafteroptic}{$0.014^{+0.015}_{-0.008}$}
\newcommand{\OfourBafterx}{$0.08^{+0.08}_{-0.04}$}
\newcommand{\OfourBafterradiodeg}{$0.10^{+0.11}_{-0.06}$}
\newcommand{\OfourBafteropticdeg}{$0.02^{+0.02}_{-0.01}$}
\newcommand{\OfourBafterxdeg}{$0.22^{+0.23}_{-0.13}$}
\newcommand{\OfourCafterradio}{$0.03^{+0.04}_{-0.02}$}
\newcommand{\OfourCafteroptic}{$0.007^{+0.007}_{-0.004}$}
\newcommand{\OfourCafterx}{$0.04^{+0.04}_{-0.02}$}
\newcommand{\OfourCafterradiodeg}{$0.04^{+0.05}_{-0.03}$}
\newcommand{\OfourCafteropticdeg}{$0.004^{+0.004}_{-0.002}$}
\newcommand{\OfourCafterxdeg}{$0.11^{+0.11}_{-0.06}$}
\newcommand{\OfourDafterradio}{$0.13^{+0.13}_{-0.07}$}
\newcommand{\OfourDafteroptic}{$0.03^{+0.03}_{-0.02}$}
\newcommand{\OfourDafterx}{$0.16^{+0.16}_{-0.09}$}
\newcommand{\OfourDafterradiodeg}{$0.21^{+0.21}_{-0.12}$}
\newcommand{\OfourDafteropticdeg}{$0.03^{+0.03}_{-0.02}$}
\newcommand{\OfourDafterxdeg}{$0.55^{+0.57}_{-0.32}$}
\newcommand{\OfiveAafterradio}{$0.14^{+0.14}_{-0.08}$}
\newcommand{\OfiveAafteroptic}{$0.05^{+0.05}_{-0.03}$}
\newcommand{\OfiveAafterx}{$0.07^{+0.07}_{-0.04}$}
\newcommand{\OfiveAafterradiodeg}{$0.23^{+0.24}_{-0.14}$}
\newcommand{\OfiveAafteropticdeg}{$0.07^{+0.08}_{-0.04}$}
\newcommand{\OfiveAafterxdeg}{$0.18^{+0.19}_{-0.11}$}
\newcommand{\OfiveBafterradio}{$0.81^{+0.83}_{-0.47}$}
\newcommand{\OfiveBafteroptic}{$0.23^{+0.24}_{-0.14}$}
\newcommand{\OfiveBafterx}{$0.33^{+0.34}_{-0.20}$}
\newcommand{\OfiveBafterradiodeg}{$1.46^{+1.50}_{-0.85}$}
\newcommand{\OfiveBafteropticdeg}{$0.73^{+0.75}_{-0.43}$}
\newcommand{\OfiveBafterxdeg}{$1.22^{+1.25}_{-0.71}$}
\newcommand{\OfiveCafterradio}{$0.55^{+0.56}_{-0.32}$}
\newcommand{\OfiveCafteroptic}{$0.21^{+0.21}_{-0.12}$}
\newcommand{\OfiveCafterx}{$0.27^{+0.27}_{-0.15}$}
\newcommand{\OfiveCafterradiodeg}{$0.92^{+0.95}_{-0.54}$}
\newcommand{\OfiveCafteropticdeg}{$0.42^{+0.43}_{-0.25}$}
\newcommand{\OfiveCafterxdeg}{$0.92^{+0.94}_{-0.53}$}
\newcommand{\OfiveDafterradio}{$1.9^{+2.0}_{-1.1}$}
\newcommand{\OfiveDafteroptic}{$0.47^{+0.49}_{-0.28}$}
\newcommand{\OfiveDafterx}{$0.71^{+0.73}_{-0.41}$}
\newcommand{\OfiveDafterradiodeg}{$3.2^{+3.3}_{-1.9}$}
\newcommand{\OfiveDafteropticdeg}{$1.7^{+1.8}_{-1.0}$}
\newcommand{\OfiveDafterxdeg}{$2.8^{+2.9}_{-1.6}$}

\newcommand{\OfourApromptfermi}{$0.012^{+0.013}_{-0.007}$}
\newcommand{\OfourApromptswift}{$0.002^{+0.002}_{-0.001}$}
\newcommand{\OfourApromptfermideg}{$0.044^{+0.045}_{-0.026}$}
\newcommand{\OfourApromptswiftdeg}{$0.007^{+0.007}_{-0.004}$}
\newcommand{\OfourBpromptfermi}{$0.04^{+0.04}_{-0.02}$}
\newcommand{\OfourBpromptswift}{$0.006^{+0.006}_{-0.004}$}
\newcommand{\OfourBpromptfermideg}{$0.20^{+0.21}_{-0.12}$}
\newcommand{\OfourBpromptswiftdeg}{$0.04^{+0.04}_{-0.02}$}
\newcommand{\OfourCpromptfermi}{$0.02^{+0.02}_{-0.01}$}
\newcommand{\OfourCpromptswift}{$0.003^{+0.003}_{-0.002}$}
\newcommand{\OfourCpromptfermideg}{$0.12^{+0.12}_{-0.07}$}
\newcommand{\OfourCpromptswiftdeg}{$0.2^{+0.02}_{-0.01}$}
\newcommand{\OfourDpromptfermi}{$0.07^{+0.07}_{-0.04}$}
\newcommand{\OfourDpromptswift}{$0.012^{+0.012}_{-0.007}$}
\newcommand{\OfourDpromptfermideg}{$0.5^{+0.5}_{-0.3}$}
\newcommand{\OfourDpromptswiftdeg}{$0.09^{+0.09}_{-0.05}$}
\newcommand{\OfiveApromptfermi}{$0.04^{+0.04}_{-0.02}$}
\newcommand{\OfiveApromptswift}{$0.008^{+0.008}_{-0.005}$}
\newcommand{\OfiveApromptfermideg}{$0.15^{+0.15}_{-0.09}$}
\newcommand{\OfiveApromptswiftdeg}{$0.03^{+0.03}_{-0.01}$}
\newcommand{\OfiveBpromptfermi}{$0.20^{+0.20}_{-0.11}$}
\newcommand{\OfiveBpromptswift}{$0.04^{+0.05}_{-0.02}$}
\newcommand{\OfiveBpromptfermideg}{$0.92^{+0.94}_{-0.54}$}
\newcommand{\OfiveBpromptswiftdeg}{$0.17^{+0.19}_{-0.09}$}
\newcommand{\OfiveCpromptfermi}{$0.17^{+0.17}_{-0.10}$}
\newcommand{\OfiveCpromptswift}{$0.03^{+0.03}_{-0.02}$}
\newcommand{\OfiveCpromptfermideg}{$0.64^{+0.65}_{-0.37}$}
\newcommand{\OfiveCpromptswiftdeg}{$0.12^{+0.16}_{-0.06}$}
\newcommand{\OfiveDpromptfermi}{$0.4^{+0.4}_{-0.2}$}
\newcommand{\OfiveDpromptswift}{$0.07^{+0.08}_{-0.04}$}
\newcommand{\OfiveDpromptfermideg}{$2.2^{+2.2}_{-1.3}$}
\newcommand{\OfiveDpromptswiftdeg}{$0.4^{+0.5}_{-0.2}$}

\newcommand{\fracOfourAem}{2.2\%}

   \title{Multi-messenger prospects for black hole -- neutron star mergers in the O4 and O5 runs\thanks{The data produced in this work are publicly available on Zenodo through the link \url{https://zenodo.org/doi/10.5281/zenodo.10700748}. The scripts and files to reproduce the main figures in the text are publicly available at \url{https://github.com/acolombo140/O4O5BHNS}.}}
   \authorrunning{A. Colombo et al.}
   \titlerunning{Multi-messenger prospects for black hole -- neutron star mergers in the O4 and O5 runs}

   \author{Alberto Colombo\inst{1, 2, 3}, Raphaël Duqué\inst{4}, Om Sharan Salafia\inst{3,2}, Floor S. Broekgaarden\inst{5,6,7,8,9}, Francesco Iacovelli\inst{10,11}, Michele Mancarella\inst{1,2}, Igor Andreoni
\inst{12,13,14}, Francesco Gabrielli\inst{15}, Fabio Ragosta\inst{16,17}, Giancarlo Ghirlanda \inst{3,2}, Tassos Fragos\inst{18,11}, Andrew J. Levan \inst{19,20}, Silvia Piranomonte\inst{16}, Andrea Melandri\inst{16,3}, Bruno Giacomazzo\inst{1, 2, 3} and Monica Colpi\inst{1,2,3}}

   \institute{Università degli Studi di Milano-Bicocca,
              Dip.\ di Fisica "G. Occhialini", piazza della Scienza 3, I-20126 Milano (MI), Italy
         \and
             INFN -- sezione di Milano-Bicocca, Piazza della Scienza 3, I-20126 Milano (MI), Italy
         \and
             INAF -- Osservatorio Astronomico di Brera, via Emilio Bianchi 46, I-23807 Merate (LC), Italy
        \and
            Institut für Theoretische Physik, Goethe Universität Frankfurt am Main, D-60323 Frankfurt am Main, Germany
        \and 
            {Simons Society of Fellows, Simons Foundation, New York, NY 10010, USA}
        \and 
            {Department of Astronomy and Columbia Astrophysics Laboratory, Columbia University, 550 W 120th St, New York, NY 10027, USA}
        \and    
            {William H. Miller III Department of Physics and Astronomy, Johns Hopkins University, Baltimore, Maryland 21218, USA}
        \and 
            {Center for Astrophysics \textbar{} Harvard \& Smithsonian,
                60 Garden St., Cambridge, MA 02138, USA}
         \and 
            {AstroAI at the Center for Astrophysics \textbar{} Harvard \& Smithsonian,
                60 Garden St., Cambridge, MA 02138, USA}
        \and 
            D\'epartement de Physique Th\'eorique, Universit\'e de Gen\`eve, 24 quai Ernest Ansermet, 1211 Gen\`eve 4, Switzerland
        \and
            Gravitational Wave Science Center (GWSC), Universit\'e de Gen\`eve, 24 quai E. Ansermet, CH-1211 Geneva, Switzerland
        \and
            Joint Space-Science Institute, University of Maryland, College Park, MD 20742, USA
        \and
            Department of Astronomy, University of Maryland, College Park, MD 20742, USA
        \and
            Astrophysics Science Division, NASA Goddard Space Flight Center, Mail Code 661, Greenbelt, MD 20771, USA
            \and
            SISSA, via Bonomea 265, 34136 Trieste, Italy
            \and
            INAF -- Osservatorio Astronomico di Roma, via Frascati 33, I-00078 Monte Porzio Catone (RM), Italy
            \and
            Space Science Data Center - Agenzia Spaziale Italiana, via del Politecnico, s.n.c, 00133 Roma, Italy
        \and
            Département d'Astronomie, Université de Genève, Chemin Pegasi 51, CH-1290 Versoix, Switzerland
        \and
            Department of Astrophysics/IMAPP, Radboud University, 6525 AJ Nijmegen, The Netherlands
        \and
            Department of Physics, University of Warwick, Coventry, CV4 7AL, UK
            }

  \abstract{
The existence of merging black hole-neutron star (BHNS) binaries has been ascertained through the observation of their gravitational wave (GW) signals. However, to date, no definitive electromagnetic (EM) emission has been confidently associated with these mergers. Such an association could help unravel crucial information on these systems, for example, their BH spin distribution, the equation of state (EoS) of the neutron star and the rate of heavy element production. We modeled the multi-messenger (MM) emission from BHNS mergers detectable during the fourth (O4) and fifth (O5) observing runs of the LIGO-Virgo-KAGRA (LVK) GW detector network in order to provide detailed predictions that can help enhance the effectiveness of observational efforts and extract the highest possible scientific information from such remarkable events. Our methodology is based on a population synthesis approach, which includes the modeling of the signal-to-noise ratio of the GW signal in the detectors, the GW-inferred sky localization of the source, the  kilonova (KN) optical and near-infrared light curves, the relativistic jet gamma-ray burst (GRB) prompt emission peak photon flux, and the GRB afterglow light curves in the radio, optical, and X-ray bands. The resulting prospects for BHNS MM detections during O4 are not promising, with an LVK  GW  detection rate of \OfourALVKduty\ yr$^{-1}$, but joint MM rates of $\sim 10^{-1}$ yr$^{-1}$ for the KN and $\sim 10^{-2}$ yr$^{-1}$ for the jet-related emission. In O5, we found an overall increase in expected detection rates by around an order of magnitude, owing to both the enhanced sensitivity of the GW detector network and the coming online of future EM facilities. Considering variations in the NS EoS and BH spin distribution, we find that the detection rates can increase further by up to a factor of several tens. Finally, we discuss direct searches for the GRB radio afterglow with large field-of-view instruments during O5 and beyond as a new possible follow-up strategy in the context of ever-dimming prospects for KN detection due to the recession of the GW horizon.}


\begin{document} 

   \maketitle
%

\section{Introduction}
\label{sec:intro}

The first detections of binary compact objects composed of a black hole and a neutron star (BHNS binaries) through gravitational waves (GWs) recently accelerated the study of BHNS physics \citep{Abbott2021d}. To date, there are four BHNS events that have been detected with false alarm rate (FAR) of less than 1 $ \rm{yr}^{-1}$: GW200115\_042309, GW200105\_162426, GW190917\_114630, and GW190426\_152155
\citep{2023PhRvX..13a1048A}, by the LIGO-Virgo-KAGRA Scientific Collaboration (LVK) consisting of the two Advanced Laser Interferometer Gravitational-wave Observatories \citep[aLIGO,][]{Aasi2015}, the Advanced Virgo \citep{Acernese2015}, and KAGRA detector \citep{KAGRA}.

BHNS observations provide novel and complementary information on the formation pathways of compact objects to be compared with those of binary black hole (BBH) and binary neutron star (BNS) mergers \citep[e.g., ][]{Kruckow2018, Santoliquido2021, 2022MNRAS.516.5737B, Wagg2022}. The first detections of mergers suggest that BHNS binaries are a population hosting highly asymmetric binaries with a large mismatch between the BH and NS mass compared to BBHs and BNSs, suggesting different progenitor stars and formation avenues  \citep[e.g.,][]{2022MNRAS.516.5737B, 2022MNRAS.511.1454G, 2022ApJ...936..184M}.

In a BHNS merger, the NS undergoes one of two faiths: It is either partially or completely torn apart by the tidal forces of the BH outside the innermost stable circular orbit (ISCO), or it is directly engulfed by the BH \citep[e.g.,][]{kawaguchi2015,Foucart2018, Foucart2019}. The outcome of the NS hinges on the relative position between the BH ISCO $R_\mathrm{ISCO}$ and the distance $d_\mathrm{tidal}$ at which the BH gravitational field is capable of causing tidal disruption of the star. The condition $d_\mathrm{tidal}>R_\mathrm{ISCO}$ depends on several factors, and it is favored for larger values of the NS tidal deformability, larger BH spins, and lower BH masses \citep[e.g.,][]{kawaguchi2015,Foucart2018, Barbieri2020}.
When the NS is disrupted in a BHNS merger, neutron-rich material is released. This tidal debris is comprised of two distinct components: one is a gravitationally bound portion that forms an accretion disk around the BH remnant, the other is an unbound component, commonly referred to as the "dynamical ejecta"  \citep[e.g.,][]{kawaguchi2015,Foucart2018, Foucart2019}. The presence of this matter outside the BH can potentially power a kilonova (KN) and launch a relativistic jet, which can in turn produce "prompt" and "afterglow" emission, possibly contributing a sub-class of gamma-ray bursts  \citep[GRBs; e.g.,][]{Li1998, Metzger2019, 2020ApJ...895...58G, 2023arXiv230507582G, 2023MNRAS.524.3537G}. Detecting electromagnetic (EM) emission from a BHNS merger would help in constraining the equation of state (EoS) of matter at nuclear densities \citep{2012PhRvD..85d4061L,2010CQGra..27k4106D}, measuring the Hubble constant \citep[e.g.,][]{1986Natur.323..310S,2013arXiv1307.2638N,2021PhRvL.126q1102F, 2023arXiv230802440F}, and understanding the role of BHNS mergers in the production of heavy elements \citep{1974ApJ...192L.145L,1976ApJ...210..549L,1999A&A...341..499R,2011ApJ...738L..32G,2015MNRAS.448..541J}.

When compared with expectations from numerical relativity, the signals from the two BHNSs detected in January 2020 suggest that these mergers did not lead to significant ejection of matter \citep[][and references therein]{Abbott2021d, Biscoveanu2023}. This is consistent with the non-detection of EM signals during the follow-up campaign \citep[e.g.,][]{2021NatAs...5...46A}\footnote{A summary of the follow-up campaigns of these two events can be evinced from the Global Coordinates Network circulars at \url{https://gcn.gsfc.nasa.gov/other/S200105ae.gcn3}, and \url{https://gcn.gsfc.nasa.gov/other/S200115j.gcn3}.}, although the uncertainties in the distance and GW sky localization region disallow completely excluding the possibility of a potentially detectable EM emission. 
Moreover, the fraction of BHNS bright events is commonly considered low \citep{Fragione2021}, as the possibility of highly rotating BHs and stiff NS EoS is disfavored by LVK constraints \citep{Abbott2021gwtc3,Abbott2018}. However, the possibility of BHNS mergers with BHs in the lower mass gap ($M_\mathrm{BH}<5M_\odot$) would allow for EM counterparts even under the assumption of non-rotating BHs and soft EoSs. In particular, the hypothesis of events in the mass gap is suggested both by LVK observations \citep{2020ApJ...896L..44A,2020ApJ...899L...1Z} and by "delayed" supernova model theories \citep{fryer2012,drozda2022,Broekgaarden21}, upon which the population considered in this study is based.
The current and upcoming observation runs of the global network of GW interferometers (IFOs) with an ever-improving sensitivity open the doors to elucidating the elusive properties of BHNS systems, detecting their GW signals and, for a subclass of these, their joint GW and EM signals.

This study is an attempt to infer realistic prospects of the rate of BHNS mergers as multi-messenger (MM) sources  during the current fourth (O4) and upcoming fifth (O5) runs and to explore their properties. Our study is based on population synthesis models for the BHNS systems, numerical relativity-informed prescriptions for the properties of the materials expelled from the mergers, and semi-analytical models to compute the observable properties of the associated KN, GRB prompt, and GRB afterglow emission.

Before run O4 started (in May 2023), a number of papers appeared in the literature that anticipated the estimates of the MM detection rates of BHNS mergers. \citet{Boersma2022} focused on the observation of GRB radio afterglows with the SKA1 radio array, finding that a joint detection is unlikely within O5, even with such a sensitive instrument. They pointed out how current uncertainties on the BH spin greatly affect the results, as expected from the role that the BH spin plays in the disruption of the NS. Similarly, \citet{2021ApJ...917...24Z} focused on the KN and GRB afterglow counterparts of BHNS mergers and found a prominence of plunging systems with negligible EM emission for a population of low-spinning BHs. Consequently, they suggested relying on searches in data streams from optical surveys as a strategy to discover KN and afterglow counterparts of BHNS systems (so-called blind searches, e.g., \citealt{2020ApJ...904..155A,2021ApJ...918...63A}) rather than dedicating target-of-opportunity (ToO) time to EM facilities after GW triggers with low chances of a successful outcome. Regarding the GRB prompt counterpart, \citet{Zhu2022} was the first to derive a population model for BHNS systems starting from three long-duration GRBs (including GRB211211A), under the hypothesis of a BHNS origin. They estimated a joint GW+KN+GRB detection rate of BHNS mergers of $\sim0.1\:{\rm yr}^{-1}$ during O4, though large uncertainties remain.

In this study, we extend the analysis of the MM properties of BHNS mergers considering the GRB prompt, multi-wavelength GRB afterglow, and KN emissions for BHNS systems detectable above the GW network threshold. Furthermore, we study the impact of the EoS of NS matter and of the BH spin distribution on the expected population of MM events. This enables us to account for two of the main sources of uncertainty among BHNS systems. Also, our predictions for detectable EM counterparts are based on the follow-up performance expected for existing and planned instruments across the EM spectrum. Finally, our predictions for the MM signals allow us to derive follow-up strategies tailored to BHNS events.

The paper is organized as follows. In Sec.~\ref{sec:popmodel}, we describe the distribution of binary parameters in our BHNS population model, the emission models for the GW and EM signals, and the assumed MM representative detection limits for O4 and O5. In Sec.~\ref{sec:mm}, we present our results for the MM detection rates, and in Sec.~\ref{sec:detailed} we detail the expected properties of the different EM counterparts. We discuss these results in Sec.~\ref{sec:discussion}, specifically comparing our results with similar studies in the literature. We also discuss observing strategies that would best target the BHNS mergers. Finally, in Sec.~\ref{sec:conclusion} we summarize our results and conclude the paper.Throughout this work, we assume a flat cosmology with cosmological parameters drawn from \citet{Planck2020}. 

\section{GW and EM population models}
\label{sec:popmodel}

\subsection{Progenitor binary population}
\label{sec:channels}

\begin{figure}[t]
    \centering
    \includegraphics[width=\columnwidth]{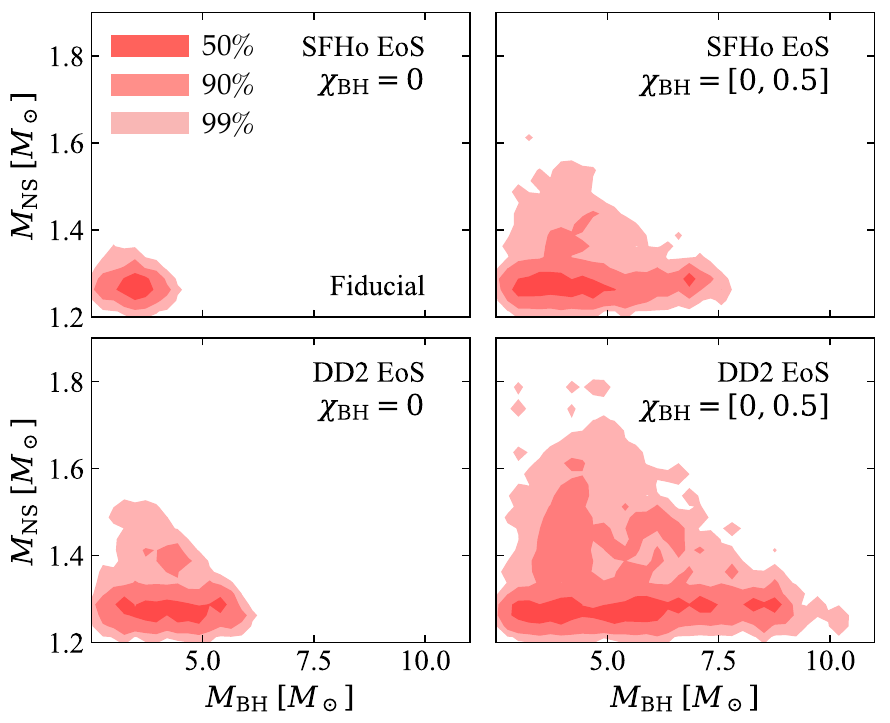}
    \caption{Distribution of BHNS binaries in our population on the NS mass versus BH mass plane, restricted to events for which the mass $m_\mathrm{out}$ remaining outside the remnant is larger than zero \cite[][Eq.\ 9]{Kruger2020}, the fundamental condition for an EM emission. In each panel, the red shaded regions contain $50\%$, $90\%$ and $99\%$ of the binaries (darker to lighter shades). Upper panels assume the SFHo EoS, while lower panels assume the DD2 EoS. Left-hand panels assume all BH spins are negligible, while right-hand panels assume uniformly distributed BH spin parameters in the interval [0,0.5].}
    \label{fig:mass}
\end{figure}

Given the scarcity of observational constraints, we chose to build our population based on binary population synthesis models. In particular, we assumed the BH and NS mass distributions resulting from the fiducial parameter set (model A) from \citet{Broekgaarden21}. We assumed the fiducial metallicity specific star formation rate density from the same work, based on the phenomenological model described in \citet{Neijssel2019}, and normalized to an observationally derived merger rate density of $R_0 = 149^{+153}_{-87} \rm Gpc^{-3} \rm yr^{-1}$ at redshift $z = 0$. We computed this value to self-consistently reproduce four BHNS events detected with FAR $\leq 1$ yr$^{-1}$ \citep{Abbott2021gwtc3}, following a similar procedure to the one described in Appendix A.3 of \citet{Colombo2022}. 
For the BH spin parameter $\chi_\mathrm{BH}$ prior to the merger, we considered two different configurations: a conservative one with $\chi_\mathrm{BH} = 0$ for all binaries and a more optimistic one with a uniform distribution in the interval $\chi_\mathrm{BH} \in [0,0.5]$, which corresponds to the typical spin range found in several simulations \citep{Fuller2019,Belczynski2020,RomanGarza2021,Bavera2020,Bavera2021,Bavera2023}.

In order to explore the dependence of our results on the uncertain NS EoS, we computed the NS tidal deformabilities assuming two EoS models: the soft SFHo EoS \citep[][]{Hempel2012}, with a maximum non-rotating NS mass $M_\mathrm{TOV} = 2.06\,\mathrm{M_\odot}$ and a 1.4 $\mathrm{M_\odot}$ NS radius $R_{1.4}=11.30\,\mathrm{km}$; and the stiff DD2 EoS \citep[][]{Steiner2013}, with $M_\mathrm{TOV}=2.46\,\mathrm{M_\odot}$ and $R_{1.4}=13.25\,\mathrm{km}$. Both EoS are in agreement with current constraints from GW170817 and the Neutron star Interior Composition Explorer \citep[NICER, e.g.][]{Miller2019, Miller2021, Raaijmakers2021}, with a slight tension for the DD2, which we therefore consider as optimistic regarding the emission of EM radiation. Hereafter, we take the conservative set-up with non-spinning BHs and the SFHo EoS as our fiducial population and consider the other pairs of BH spin distributions and NS EoS as variations on this fiducial model. 

In Figure~\ref{fig:mass} we show, for all the four variations, the regions containing $50\%$, $90\%$, and $99\%$ of the binaries with a mass remaining outside the remnant  $m_\mathrm{out} >0 $, which is the fundamental condition for having an EM emission. Here, we determine this mass using numerical-relativity-informed fitting functions from \citet[][Eq.\ 9]{Kruger2020}, see section \ref{sec:EMmodel}. Our fiducial scenario, which assumes no spins for the BH and a soft EoS for the NS, is therefore the most pessimistic scenario in terms of EM emission. Indeed, in the fiducial model, only events with a $M_\mathrm{BH} \lesssim 4.5 M_\odot$ can emit an EM counterpart. By varying the spin and the EoS with more optimistic assumptions, it is possible to find EM counterparts associated with events with a $M_\mathrm{BH} \lesssim 11 M_\odot$. These variations also correspond to a different fraction of BHNS events with $m_\mathrm{out}>0$. We find a range between $2.2\% - 13\%$ of GW triggers that satisfied the previous condition, in agreement with different estimates in the literature \citep{RomanGarza2021,Biscoveanu2023}.

Figure~\ref{fig:mass} shows that in our fiducial population model (upper-left panel), all the BHNS systems susceptible to emit EM radiation have a small-mass BH, with $M_{\rm BH} < 5\,M_\odot$, due to their zero spins. Until the first discoveries of GW from merging compact objects, the existence of BH with such low masses was largely doubted. Indeed, the observation of galactic X-ray binaries suggested a sharp cutoff of BH masses below around 5 $M_\odot$ \citep{2010ApJ...725.1918O,2011ApJ...741..103F}, leading to the existence of a "lower mass gap" in BHs masses between $5\,M_{\odot}$ and the most massive NS at $M_{\rm TOV} \sim 2.2 - 2.5\,M_{\odot}$ (according to EoS analysis and various observational constraints; for a review, see \citealt{2016ARA&A..54..401O}). In addition, this mass gap had theoretical support from binary evolution scenarios such as the "rapid" core-collapse supernova mechanism \citep{2012ApJ...749...91F,2012ApJ...757...91B}.

However, recent GW observations could suggest the lower mass gap could not be as empty as anticipated, through the detection of the merger of a system with a secondary component likely in the bounds of the purported lower mass gap \citep{2020ApJ...896L..44A,2020ApJ...899L...1Z}. A mass distribution overlapping with the mass gap, such as ours, is supported in theory by the "delayed" supernova explosion model from which our population is drawn \citep{Broekgaarden21}.

Future GW observations will continue to test the robustness of the lower mass gap. If these observations do not support the lower mass gap, then BHNS systems such as in our fiducial model, with low-mass and non-spinning BHs, would be a viable channel for EM radiation from BHNS mergers, in addition to the high-mass, high-spin systems identified previously and included in our population model variations. In this case, the requirements on the SN scenario to produce EM-bright BHNS mergers would not be as stringent as anticipated by \citet{2023arXiv230909600X}. In any case, including a population of systems with BHs in the mass gap, such as our fiducial model, is a novelty of our work and contributes to shifting the focus from high-spin BHs to low-mass BHs as progenitors for EM-bright BHNS mergers.

Finally, we checked that our prescription for the binary masses are consistent with the current (yet weak) constraints deduced from the first observations of BHNS mergers. To this effect, as reported in Appendix~\ref{sec:mass_comparison}, we compared the mass distributions of the NS and BH component of the systems in our BHNS population model with the constraints derived by \citealt{Biscoveanu2023} from the first detections of mergers in the GW domain. We find that, indeed, our prescriptions are consistent with the first constraints deduced from GW data, though these remain quite uncertain.

\subsection{GW model}\label{sec:gwmodel}

For each merger event, as a detection statistic, we computed the expected network matched-filter signal-to-noise ratio defined as (see e.g. Chap.~7 of \citealt{Maggiore:2007ulw} for a formal derivation) 
\begin{equation}
    ({\rm S/N})^2 = \sum\nolimits_{i} ({\rm S/N})^2_i\,, \quad ({\rm S/N})^2_i = 4 \int_{f_{\rm min}}^{f_{\rm cut}} \dfrac{|\tilde{h}_{(i)}(f)|^2}{S_{n,i}(f)} {\rm d}f \,,
\end{equation}
with the index $i$ running over the detectors in the network, $f_{\rm min}=10~{\rm Hz}$, $f_{\rm cut}$ being a cutoff frequency determined by the events' parameters, $\tilde{h}_{(i)}(f)$ denoting the Fourier-domain GW strain projected onto the detector $i$ and $S_{n,i}(f)$ the $i^{\rm th}$ IFO's noise power spectral density. We also computed the 90\% credible sky localization area $\Delta \Omega_{90\%}$ for each signal. For O5, we assumed a network consisting of the two aLIGO, Advanced Virgo, KAGRA and LIGO-India (LVKI) with the projected O5 sensitivities~\citet{2020LRR....23....3A}\footnote{\label{footnote:ASDurl}The projected noise amplitude spectral densities we used can be found at \url{https://dcc.ligo.org/LIGO-T2000012/public}. For KAGRA we considered a target sensitivity of 10\,Mpc in O4 and 127\,Mpc in O5. For the other detectors we assumed the highest target sensitivity.}; for O4 we did not consider LIGO-India (hence we assumed an LVK network) and we assumed the O4 sensitivities from~\citet{2020LRR....23....3A}\textsuperscript{\ref{footnote:ASDurl}}. Both for the O4 and O5 scenarios, we performed the analyses with the inclusion of a $70\%$ uncorrelated duty cycle for each detector (this is the same value adopted in \citealt{2020LRR....23....3A}). The S/N and sky localization area computations were carried out through the \texttt{GWFAST} software package \citep{2022ApJS..263....2I,2022ApJ...941..208I}, using the \texttt{IMRPhenomNSBH} waveform model \citep{2015PhRvD..92h4050P,2019PhRvD.100d4003D} which, in \texttt{GWFAST}, depends on the detector-frame chirp mass, the mass ratio, the dimensionless spins of the two binary components projected along the orbital angular momentum, the luminosity distance, the sky position, %
the binary inclination angle with respect to the line of sight, the polarization angle,  the time of coalescence,  the phase at coalescence and the tidal deformability of the neutron star \citep{2022ApJS..263....2I}. For the parameters not discussed in Sec.\ \ref{sec:channels}, we draw the values from uninformative priors limited to their relevant physical range. We refer to \citet{2022ApJS..263....2I} for details on the prior ranges and definitions. 
\texttt{GWFAST} computes forecasts for the statistical uncertainty on the measurement of the source parameters resorting to the Fisher matrix approximation, which is valid in the high S/N limit \citep{2008PhRvD..77d2001V}. The full likelihood is approximated by a multivariate Gaussian in the parameters, with covariance equal to the inverse of the Fisher matrix. The error on the sky localization is computed as \citep{Barack:2003fp}
\begin{equation}
    \Delta\Omega_{{\rm 90}\%} = 2\pi\, \ln{10}\, |\sin\theta|\sqrt{C_{\theta\theta}\, C_{\phi\phi} - C_{\theta\phi}^2},
\end{equation}
where $\theta$ and $\phi$ here represent the polar and azimuthal angles of the source position in the sky in a geocentric frame (they are related to the right ascension and declination as ${\rm RA}=\phi$ and ${\rm dec}=\pi/2-\theta$, respectively), and $C_{\theta\theta}$, $C_{\theta\phi}$ and $C_{\phi\phi}$ are the relevant elements of the covariance matrix as estimated from the Fisher information matrix.

To keep the inversion error of the Fisher matrix under control in the subspace $\{\theta, \phi \}$, we resort to a singular value decomposition (SVD) and eliminate from the inversion singular values below a threshold value of $10^{-10}$ \citep{Dupletsa:2022scg}. This allows in particular to avoid numerical instabilities in presence of strong correlations between distance and inclination. This is particularly relevant for GW events with an associated GRB emission, which are usually close to face-on. We refer to Sec.~2 and App.~D of \citet{2022ApJ...941..208I} for a detailed discussion of this issue and for a comparison of different inversion methods. 

\subsection{EM emission models}
\label{sec:EMmodel}

For all events in our population we computed the expected mass in dynamical ejecta $m_\mathrm{dyn}$ and the average velocity $v_\mathrm{dyn}$, and the mass $m_\mathrm{out}$ remaining outside the remnant BH, using the numerical relativity-informed fitting formulae from \citet[][their Eq.\ 9]{Kruger2020}, \citet[][their Eq.\ 1]{Kawaguchi2016} and \citet[][their Eq.\ 4]{Foucart2018}. In cases where $m_\mathrm{dyn}>0.5 m_\mathrm{out}$ as predicted by these formulae, we imposed $m_\mathrm{dyn} = m_\mathrm{dyn}^\mathrm{max}(m_\mathrm{out}) = 0.5 m_\mathrm{out}$ \citep{Foucart2019,Rees1988}. Finally, we computed the mass of the accretion disk as $M_\mathrm{disk} = m_\mathrm{out} - m_\mathrm{dyn}$. 

We utilized the obtained outcomes as inputs to calculate the observable properties of the EM counterparts associated with each binary in our population, following a procedure analogous to \citet{Colombo2022} and \citet{Barbieri2019, Barbieri2020}. In particular, we computed the KN light curves in the $g$ (484 nm central wavelength), $z$ (900 nm), and $J$ (1250 nm) bands, employing the anisotropic multi-component model presented in \citet{Breschi2021}, based on \citet{Perego2017}. For events with $M_\mathrm{disk}>0$, we assumed the system to launch a relativistic jet whose energy $E_\mathrm{c}$ was computed following the method described in \citet{Colombo2022}. This method relies only on estimates of the post-merger physical quantities of remnant and disk mass and is applicable to both BHNS and BNS systems. If the ratio of jet energy to ejecta mass surpassed the threshold based on \citet[their Eq. 20]{Duffell2018}, we assumed that the relativistic jet successfully breaks out from the ejecta cloud, leading to the production of both GRB prompt and afterglow emissions. 

In order to compute the observables associated with the relativistic jet, the jet's angular structure has to be specified, i.e. the angular profile of the jet isotropic-equivalent energy $E(\theta)$ and bulk Lorentz factor $\Gamma(\theta)$, as a function of the latitude $\theta$ of the material from the jet axis \citep[e.g.][]{Salafia2022}. We considered two variations. In the first, we assumed the same structure used in \citet[][see their Appendix B for more details]{Colombo2022}, inspired by the GRB~170817A structure as inferred by \citet{Ghirlanda2019}. 
It features a uniform jet core of half-opening angle $\theta_j=3.4^\circ$ outside of which the energy falls off as a power law $E\propto\theta^{-5.5}$ and the bulk Lorentz factor as $\Gamma\propto\theta^{-3.5}$. The bulk Lorentz factor in the core was fixed at $\Gamma_\mathrm{c}=250$. The core isotropic-equivalent energy was set based on the requirement that the total jet energy equals a fraction of the accretion disk rest-mass energy, following the method described in \citet[][see also \citealt{Barbieri2019}]{Colombo2022}. In brief, such fraction is $\sim 10^{-3}$ if the remnant BH spin is $a_\mathrm{BH,rem}\sim 0.7$ \citep[based on the accretion-to-jet energy conversion efficiency of GW170817 as estimated in][]{Salafia2021}, and the dependence on $a_\mathrm{BH,rem}$ is based on the \citealt{Blandford1977} mechanism efficiency as derived in \citet{Tchekhovskoy2010}.  

In order to compute the photon flux in the gamma-ray detector band, an additional assumption on the spectrum is needed. We obtained it by assuming a latitude-independent comoving spectral shape that is a power law with an exponential cut-off, with a comoving $\nu F_\nu$ peak photon energy $E_\mathrm{peak}^\prime=3.2\,\mathrm{keV}$ (similar to \citealt{Salafia2019} and identical to \citealt{Colombo2022}), and summing the contributions from all jet latitudes after accounting for relativistic beaming, as done in \citet{Salafia2015}. We defer the reader to Appendix B of \citet{Colombo2022} for more details on the computation of the prompt emission properties.

While the choice of jet structure facilitates the comparison with our BNS results from \citet{Colombo2022}, the typical jet structure of BNS and BHNS jets could arguably differ, due to the distinct environments in which the corresponding jets are launched. In particular, the jet self-collimation due to the development of a hot, over-pressured cocoon  may be less effective owing to the presumably lower density in the polar region of the BHNS post-merger system, with respect to a BNS system \citep{Bromberg2011,Duffell2015,Lazzati2019,Urrutia2021,Salafia2020,Hamidani2021,Gottlieb2022,Salafia2022}. Indeed, in the latter, shock-driven ejecta arising from the collision of the two stars are expected to lead to a more isotropic distribution of dynamical ejecta compared to to the BHNS case \citep{Foucart2020}, in which dynamical ejecta are primarily produced near the equatorial plane due to tidal disruption of the NS \citep{Kawaguchi2016}. In addition, the intermediate supra- or hyper-massive NS state in BNS mergers, which is expected to generate strong post-merger winds \citep[e.g.][]{Fernandez2013,Just2015}, is absent in the BHNS case. For these reasons, we also considered a larger jet half-opening angle $\theta_j=15^\circ$, which is our second variation for the choice of the jet structure. In this case, we left all the other jet structure parameters unchanged, with the exception of the jet core isotropic-equivalent energy $E_\mathrm{c}=E(0)$, which was rescaled to keep the total jet energy constant. In Appendix \ref{sec:appB} we show the dependence of the prompt properties on the viewing angle $\theta_\mathrm{v}$ for the two assumed values of $\theta_j$.

For each jet structure, we generated afterglow light curves spanning from $0.1$ to $1000$ days in the radio ($1.4$ GHz), optical ($g$ band), and X-rays (1 keV) assuming a fixed interstellar medium number density of $n=5\times10^{-3},\mathrm{cm^{-3}}$ (the median density in the \citealt{Fong2015} sample) and employing afterglow microphysical parameters of $\epsilon_\mathrm{e}=0.1$, $\epsilon_\mathrm{B}=10^{-3.9}$, and $p=2.15$, representative values for GW170817, as reported in \citet{Ghirlanda2019}. These physical parameters pertain to the microphysical behavior in the shock system formed when the jet decelerates in the circum-burst medium. Therefore, they are set solely by the jet's energy and Lorentz factor, which we have chosen to be as inferred in GW170817, hence our choice of microphysical parameters as in GW170817. Concerning the GRB prompt, we employed a semi-phenomenological model, similar to the approach utilized in previous studies such as \citet{Barbieri2019} and \citet{Salafia2019}. This model assumes that a constant fraction $\eta_\gamma=0.15$ \citep{Beniamini2016} of the jet energy density, limited to regions with a bulk Lorentz factor $\Gamma\geq 10$, is radiated in the form of photons with a fixed spectrum in the jet comoving frame \citep[Sec. B.3]{Colombo2022}. The observed spectrum was then derived by integrating the resulting radiation across the solid angle of the jet, accounting for relativistic beaming at the relevant viewing angle.

\subsection{Multi-messenger detection criteria}
\label{sec:limits}

The sub-population of BHNS systems that will be detectable and that can provide MM datasets is determined by the instruments available for the GW observations and the follow-up efforts. 

For the detection of the GW signals in both O4 and O5, we applied a network S/N threshold of 12. This limit is the same as that assumed in our previous study of BNS systems \citep{Colombo2022} and it is representative of the S/N threshold for a confident detection by the LVK Collaboration \citep{2020LRR....23....3A}. Such relatively high S/N cut also makes the \texttt{GWFAST} Fisher-information-matrix-based parameter estimation forecast more reliable \citep{2022ApJ...941..208I}.

For EM follow-up during O4, we adopted the same limits in the radio, optical, X-ray and gamma-ray bands as in \citet{Colombo2022}, that is, a limiting radio flux density of $0.1\,{\rm mJy}$ at $1.4\,{\rm GHz}$, representative of the limits for current radio arrays adopting the "galaxy targeted" or "unbiased" search strategies \citep[e.g.][]{2021MNRAS.505.2647D}; optical and near-infrared limiting AB magnitudes of $g<22$, $z<22$ and $J<21$ respectively, in line with the typical depths reached in EM counterpart searches during the O3 run \citep[e.g.,][]{Coughlin2019,2020A&A...643A.113A, 2020ApJ...890..131A} and with new wide-field instruments \citep{De2020,Lourie2020}; X-ray limiting flux of $10^{-13}\,{\rm erg/cm}^2{\rm /s/keV}$ at 1\,keV, representative of the limits that can be reached by Chandra or XMM-Newton with long exposures ($\sim 10^4\,{\rm s}$, e.g., \citealt{2017ApJ...848L..20M}; \citealt{2018A&A...613L...1D}); in the gamma-ray band for the detection of the GRB prompt emission, a limiting 10-1000\,keV average photon flux of $4\,{\rm ph\, cm^{-2}\, s^{-1}}$, as deduced from the cumulative distribution of photon fluxes of GRBs in \textit{Fermi}/GBM which is our reference instrument \citep[][Appendix B.3]{Colombo2022}.

From O4 to O5, we expect a significant improvement in the depth of the search for counterparts in the radio and optical bands, owing to new instruments coming online. In the radio band, we considered an order of magnitude improvement with a limiting flux of $0.01\,{\rm mJy}$ \citep{2021MNRAS.505.2647D}. We note that this flux level was approximately that of the afterglow in GW170817 upon discovery in the $3-6\,{\rm GHz}$ bands in deep searches at the location of the KN transient \citep{2017ApJ...848L..12A,2021ApJ...922..154M}, but this level could be reached by the next-generation instruments in untargeted counterpart searches, such as the SKA2 \citep{2019arXiv191212699B}, Next-Generation VLA \citep{2019arXiv190310589C} or DSA-2000 \citep{2019BAAS...51g.255H}. 

In the optical band, we accounted for the arrival of new large field of view (FoV) instruments such as the Vera Rubin Observatory \citep[aka, "the Rubin Observatory"][]{2008SerAJ.176....1I}. For this purpose, we considered a magnitude threshold of 26 in the $g$ band and 24.4 in the $z$ band, corresponding to the limit reached in the preferred follow-up strategy suggested for the target-of-opportunity program of Rubin Observatory \citep[180\,s exposure,][]{2022ApJS..260...18A}.

However, for the X-ray and gamma-ray bands, we considered the same limits in O5 as in O4, due to the later launching of the next generation of high-energy instruments with significant improvements in both the FoV and sensitivity, such as \textit{THESEUS} \citep{2021ExA....52..183A}, \textit{HERMES} \citep{2020SPIE11444E..1RF} or the \textit{Gamow Explorer} \citep[][2030+]{2021SPIE11821E..09W}. Although we did not consider these instruments for follow-up in O4 and O5, in Sec.~\ref{sec:direct} we developed new follow-up strategies tailored to BHNS mergers inspired by recently proposed strategies for upcoming large-FoV X-ray instruments, such as \textit{ATHENA}/WFI \citep{2013arXiv1306.2307N} which could tile the GW localization sky map in search for the X-ray afterglow counterpart \citep{2022A&A...665A..97R}.

We stress that, in this work, we operated on the premise that the GW sky localization areas of O4 and O5 BHNS mergers will consistently be surveyed to the assumed detection thresholds by the combined efforts of various observatories, as evidenced by the EM counterpart searches following the BNS event GW190425 \citep{Coughlin2019,Hosseinzadeh2019,Antier2020}. A more detailed evaluation of the actual detection rates attainable in practical terms would necessitate conducting simulations that mimic the search strategies implemented by individual facilities. In Appendix \ref{sec:appC}, we explore how our prospects for the detection rates change by varying our assumptions about the detection limits and the events sky localization.

\section{Multi-messenger detection prospects}
\label{sec:mm}

\begin{figure*}
    \centering
    \includegraphics[width=\textwidth]{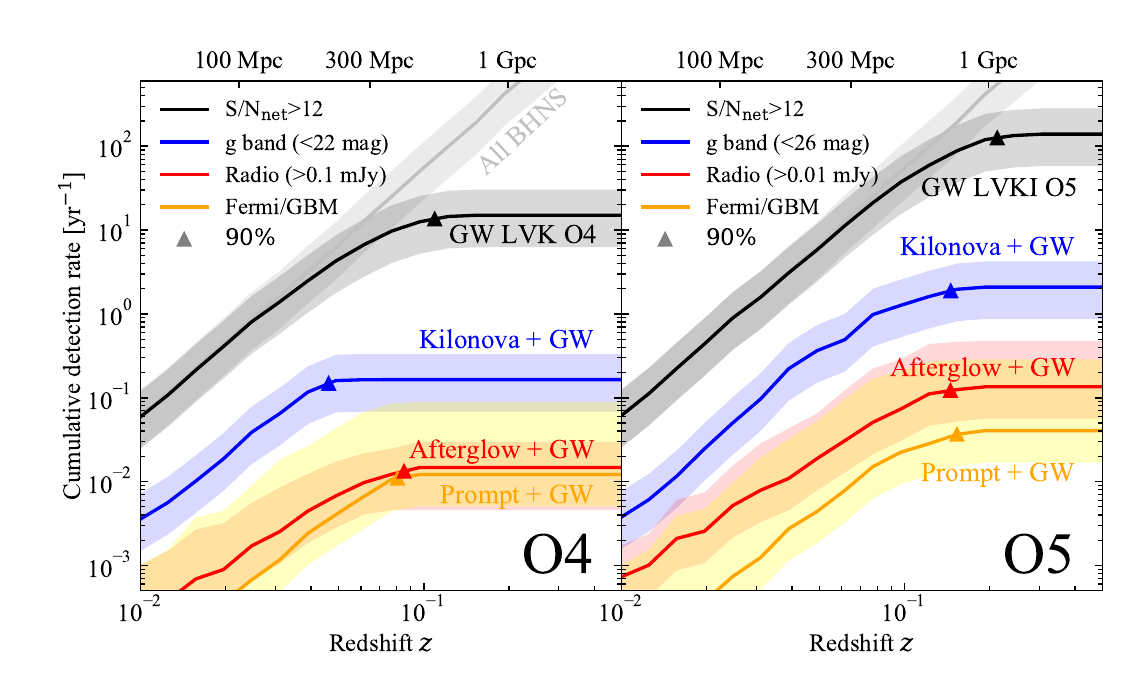}
    \caption{Cumulative MM detection rates as a function of redshift (luminosity distance) for our fiducial BHNS population (SFHo EoS, non-spinning BHs) and assuming a jet core half-opening angle $\theta_j = 3.4^{\circ}$ for the jet-related emissions. Triangles indicate the 90$^\mathrm{th}$ percentile of the cumulative detection rate (Fig.\ \ref{fig:pop_variations} shows how this is affected when varying our assumptions). \textit{Left-hand panel}: assumes the LVK network configuration and the O4 projected sensitivities. The light gray line ("All BHNS") represents the intrinsic merger rate, with the gray band showing its uncertainty drawn from that on the local merger rate. This uncertainty propagates as a constant relative error contribution to all the other rates shown in the figure. The error region for the jet-related emissions takes also into account a possible larger core half-opening angle $\theta_j = 15^{\circ}$. The black ("GW LVK O4") line is the cumulative GW detection rate (events per year with network S/N $\geq 12$) in O4. The blue ("Kilonova+GW"), red ("Afterglow+GW") and orange ("Prompt+GW") lines are the cumulative detection rates for the joint detection of GW plus either a KN, a GRB afterglow or a GRB prompt in O4 (all-sky except for the orange line, which accounts for the \textit{Fermi}/GBM duty cycle and field of view). The assumed thresholds are shown in the legend. \textit{Right-hand panel:} similar to the left-hand panel, but assuming an LVKI GW detector network with the O5 projected sensitivities, and deeper EM detection thresholds (see text).}
    \label{fig:detection_rates}
\end{figure*}

\begin{figure*}
    \centering
    \includegraphics[width=\textwidth]{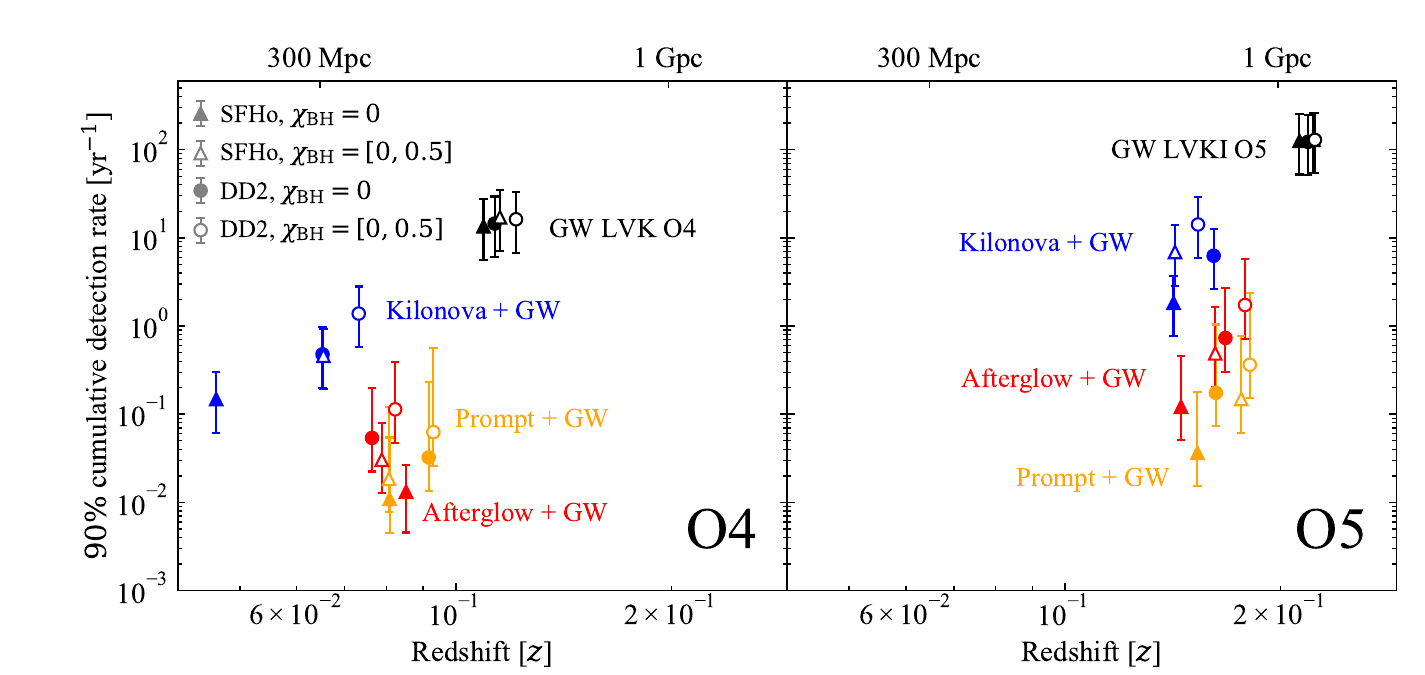}
    \caption{Predicted 90$^\mathrm{th}$ percentile of the cumulative MM detection rates for our four population model variations, assuming a jet core half-opening angle $\theta_j = 3.4^{\circ}$ for the jet-related emissions. Different marker shapes indicate different adopted EoSs (triangle: SFHo; circle: DD2). Filled markers are for $\chi_\mathrm{BH} = 0$, while empty markers are for a uniform spin parameter distribution between 0 and 0.5. The error bars indicate the uncertainty on the local merger rate. For GRB afterglow and prompt they also take into account a variation on the jet core half-opening angle ($\theta_j = 15^{\circ}$, corresponding to the minimum and maximum value reported in Table \ref{tab:det_rates}).  Similarly as in Fig.\ \ref{fig:detection_rates}, the left-hand panel assumes an LVK GW-detector network and the O4 projected sensitivities, while the right-hand panel assumes an LVKI network and the O5 sensitivities. EM bands and detection thresholds are the same as in Figure \ref{fig:detection_rates}.}
    \label{fig:pop_variations}
\end{figure*}

Using the synthetic population described in Sec.~\ref{sec:popmodel}, we can determine the detection rates of BHNS mergers in the GW domain and in the various EM counterparts by applying S/N and peak brightness cuts to each of the events in the population. We start by presenting the results for our fiducial set of assumptions and then study the variations on the EoS, BH spin distribution, and jet core opening angle.

In Fig.~\ref{fig:detection_rates}, we present cumulative detection rates as a function of redshift, with error propagated from the uncertainty on the intrinsic merger rate as described in Sec.~\ref{sec:channels} and considering the variation on the jet core opening angle $\theta_j$ for the jet related emissions (see Fig.~\ref{fig:pop_variations} and below for the uncertainty on the assumptions of our population model). 

In total, we expect a BHNS GW detection rate of \OfourALVKduty\ ${\rm yr}^{-1}$ (90\% confidence interval) for our assumed BHNS mass distribution and detection thresholds during O4.
The cumulative rate of GW-detectable sources with an EM-detectable KN\footnote{In the case of all EM+GW events, we display only one band in Fig. \ref{fig:detection_rates} for clarity. The results for the other bands considered in the study can be seen in Table \ref{tab:det_rates}.} (blue curve) follows the same trend as the distribution of GW detections up to a distance of $200\,{\rm Mpc}$, indicating that the majority of KN\ae\ in our population is luminous enough to be detected within this distance. Up to $200\,{\rm Mpc}$, we expect \fracOfourAem\ of GW triggers to give rise to a detectable KN. At $d_\mathrm{L}\gtrsim 200\,\mathrm{Mpc}$ the cumulative rate of KN\ae\ flattens out as an increasing number of KN\ae\ fail to exceed the assumed detection thresholds, with the cumulative rate saturating at \OfourAkng\ ${\rm yr}^{-1}$, about two orders of magnitude lower than the total GW detections.

The cumulative rate of GW-detectable BHNS with an afterglow counterpart detectable in the radio band (red curve) features a shallower slope with respect to the GW or GW+KN cumulative rates, owing to the strong dependence of the afterglow peak brightness on the inclination angle. The total rate of radio afterglow detection is predicted to be \OfourAafterradio\ ${\rm yr}^{-1}$ (\OfourAafterradiodeg\ ${\rm yr}^{-1}$) assuming $\theta_j = 3.4^\circ$ ($\theta_j = 15^\circ$).

Concerning the GRB prompt emission (orange curve), we find a cumulative rate of GW-detectable events with an EM-detectable counterpart that increases with redshift with the same slope as that of GW events, hence with a constant ratio between the two rates on the order of $10^3$. This is due to the fact that the prompt emission luminosity drops rapidly for viewing angles outside the jet core opening angle, such that the selection of the prompt emission boils down to simply a selection of solid angle, which does not vary appreciably with distance, on top of the GW detectability selection. The latter dominates in determining the GW+EM horizon, because the distance up to which the prompt emission is detectable for an on-axis observer largely exceeds the GW horizon.

It's important to emphasize that in the O4 scenario, we considered a full network with high target sensitivities as reported in Sec. \ref{sec:gwmodel}. However, at the time of writing, only the LIGO Hanford and Livingston detectors are operational, with a BNS range between 140 and 170 Mpc. Virgo is expected to join the network in the second part of O4, scheduled to begin on March 27, 2024, with a target sensitivity of 40-80 Mpc. KAGRA will join the run in the spring of 2024, featuring a BNS range of approximately 10 Mpc \footnote{Updates are available at \url{https://observing.docs.ligo.org/plan/}}. Assuming a network consisting of only two aLIGO detectors would decrease our detection rates by about 14\%. However, since the joint detection rates are low, even an optimistic assumption for the network does not change the general sense of our analysis. The effects of different network assumptions on the sky localization are discussed in Appendix \ref{sec:appC}.

Moving to O5, we find an overall increase in detection rates by one order of magnitude, which stems from both the better GW sensitivity and the deeper EM detection limits assumed. For the KN (blue curve), whose detection horizon is primarily determined by the EM threshold, the larger median distance of the GW-detected events is largely compensated by the expected increase in sensitivity of the optical searches, such that the KN detections track the GW detections up to a higher redshift. The GW detection rate increases up to \OfiveALVKIduty\ yr$^{-1}$. The total detection rate of KN signals in O5 reaches a promising value of \OfiveAkng\ ${\rm yr}^{-1}$, despite our conservative assumptions about the EoS and BH spin. The joint detection rate of GW along with GRB jet prompt emission (orange curve) and radio afterglow (red curve, whose horizon is mostly set by the GW sensitivity) also increases, but not as dramatically. GW-detectable systems with a detectable radio afterglow reach a total rate of \OfiveAafterradio\ ${\rm yr}^{-1}$, similar to that of GW+KN in O4.

Intriguingly, some GW-detectable BHNS mergers with a detectable afterglow feature a KN whose peak brightness is below our assumed thresholds, opening the possibility for the afterglow to be the primary counterpart to some BHNS systems in O5. We come back to this point in Sec.~\ref{sec:direct}. Even in O5, the rate of GW-detectable BHNS with a detectable GRB prompt emission remains lower than one in 10 years, making the prospects for such events not particularly promising. Under the assumption of a larger opening angle for the jet, as explored in Secs.~\ref{sec:after} and \ref{sec:grb} with $\theta_j = 15^\circ$ instead of $\theta_j = 3.4^\circ$, we would expect a larger rate of GRB prompt counterparts, up to $0.3\,{\rm yr}^{-1}$. Indeed the horizon for GRB prompt emission detection largely exceeds the GW horizon, such that GRB prompt detection is ensured for lines of sight looking into the jet.

We stress that the start dates, duration, and sensitivities projected for the O5 run are based on the best current estimates. Therefore, there is a possibility that the target sensitivities might be overestimated, as well as the inclusion of LIGO-India in the network from the beginning of the run. Removing LIGO-India from the network would decrease the detection rates by about $42\%$.

In both O4 and O5, the distance of the horizon for GW+EM events, i.e., the value of the redshift corresponding to the saturation of the curves, is smaller compared to the curve requiring just the GW detection. This is because only the lighter events, and therefore those with a smaller intrinsic S/N, are capable of emitting an EM counterpart, as highlighted in Fig. \ref{fig:mass}. Thus, the GW+EM horizon is set by the GW detection of events with a BH smaller than a certain value.

The results in Fig.~\ref{fig:detection_rates} are affected by the uncertainty in the intrinsic BHNS merger rates (and also by the choice of $\theta_j$ for the jet related emissions). Moreover, they are affected by the assumption on the BHNS formation pathway that determines the binary parameter distribution (e.g., BH mass, mass ratio, and spins) and on the NS EoS. In Fig.~\ref{fig:pop_variations}, we study the effect of this uncertainty through the variations of two assumptions of our population model. For simplicity, we only show the 90 percentile of the cumulative detection rate for GW and EM signals at the redshift where this value is reached for visualization purposes. The error bars in Fig. \ref{fig:pop_variations} are computed in the same way of Fig. \ref{fig:detection_rates}, so they take into account the uncertainty on the merger rate and, for the GRB prompt and afterglow, also the variation of $\theta_j$.

Concerning the GW detections (black symbols with black error bars), the variations of the NS EoS and $\chi_{\rm BH}$ prior induce a negligible change in detection rate and distance, in line with the marginal effect of both the component spins and tidal deformability in the inspiral signal. In O4, the variations induce a large uncertainty of one order of magnitude in KN detection rate (blue symbols), which largely surpasses the intrinsic uncertainty on the BHNS merger rate that we normalize our population to. This increase in KN detection rate follows with a large increase in redshift distance to which they are observable. This shows the crucial role of formation channel and NS properties in the MM detection prospects and, in turn, the potential to constrain these with MM data in the future. The choice of the NS EoS, with the stiffer DD2, leads to a factor of a few more KN detections than SFHo, due to the larger ejecta mass. The spin distribution plays another significant role, with higher spins favoring the disruption of the NS and a significant amount of ejecta in the post-merger phase.

For the afterglow counterpart in the radio band (red symbols), the effect of a stiffer EoS is also present due to its influence on the disk mass. One must keep in mind that, beyond the binary parameters, the afterglow is also largely determined by the microphysics parameters of the jet's forward shock and the density of the circum-merger medium. The possible variation of these is another source of uncertainty. However, we focus here on the effects of EoS and BH spin distribution, because the effects of the afterglow parameters and of the density on MM population models have already been explored for BNSs, and are expected to be similar for BHNSs \citep[e.g.,][]{2019MNRAS.488.2405G,2020A&A...639A..15D}.

Considering the O5 run, the trends for O4 are reproduced in all the counterparts. However, the role of the EoS and BH spin in determining the detection rate of KN\ae\ (blue symbols) is less important. This is due to the better optical limits we consider in O5, such that an overall larger fraction of KN\ae\ is detectable (as discussed in Fig.~\ref{fig:detection_rates}), hence a lesser sensitivity to the secondary effects.

To summarize these results, we present in Tab.~\ref{tab:det_rates} the total detection rates of BHNS mergers in the O4 and O5 runs for our fiducial population model and the three variations\footnote{We note that in Tab.~\ref{tab:det_rates} we split up the detection rates for the two jet core opening angle cases to have the same uncertainty due to the merger rate assumption. Instead in Fig. \ref{fig:detection_rates} and \ref{fig:pop_variations}, we also include in the error bands of the jet emissions the variation in the angle $\theta_j$, thus considering the maximum and minimum values reported in the table. This approach allows for an immediate visualization of the upper and lower limits for the jet emissions.}. In addition, in Appendix~\ref{sec:appC}, we briefly consider the prospects of GW sky localization of the sources to discuss the MM results presented in this section. For a more detailed discussion of how to leverage the potential for source localization, see Sec.~\ref{sec:direct}.

\def\arraystretch{1.3}%
\begin{table*}[htbp]
\tiny
\centering
\caption{Detection limits and predicted detection rates for O4 and O5, assuming our fiducial population model (SFHo EoS and $\chi_\mathrm{BH}=0$) and three variations. For the jet-related emissions, we assume a half-opening angle $\theta_j = 3.4^\circ$, while in parenthesis we report the rate assuming $\theta_j = 15^\circ$.}
\begin{tabular}{lccccccccc}
\hline
\hline
\multicolumn{1}{c}{} & \multicolumn{1}{c}{GW} & \multicolumn{3}{c}{KN + GW} & \multicolumn{3}{c}{GRB Afterglow + GW} & \multicolumn{2}{c}{GRB Prompt + GW} \\
 & ~ & \textit{J} & \textit{z} & \textit{g} & Radio & Optical & X-rays & \textit{Swift}/BAT & \textit{Fermi}/GBM \\ \hline
\textbf{LVK O4} & ~ & ~ & ~ & ~ & ~ &  &  &  & \\
Limit & 12 & 21 & 22 & 22 & 0.1 & 22 & $10^{-13}$ & $3.5$ & $4$  \\ 
Rate SFHo, $\chi_\mathrm{BH} = 0$  & \OfourALVKduty & \OfourAknJ & \OfourAknz & \OfourAkng & \OfourAafterradio & \OfourAafteroptic & \OfourAafterx & \OfourApromptswift & \OfourApromptfermi \\ 
($\theta_j = 15^\circ$) & & & & & (\OfourAafterradiodeg) & (\OfourAafteropticdeg) & (\OfourAafterxdeg) & (\OfourApromptswiftdeg) & (\OfourApromptfermideg) \\ 
Rate SFHo, $\chi_\mathrm{BH} = [0,0.5]$  & \OfourBLVKduty & \OfourBknJ & \OfourBknz & \OfourBkng & \OfourBafterradio & \OfourBafteroptic & \OfourBafterx & \OfourBpromptswift & \OfourBpromptfermi \\ 
($\theta_j = 15^\circ$) & & & & & (\OfourBafterradiodeg) & (\OfourBafteropticdeg) & (\OfourBafterxdeg) & (\OfourBpromptswiftdeg) & (\OfourBpromptfermideg) \\ 
Rate DD2, $\chi_\mathrm{BH} = 0$  & \OfourCLVKduty & \OfourCknJ & \OfourCknz & \OfourCkng & \OfourCafterradio & \OfourCafteroptic & \OfourCafterx & \OfourCpromptswift & \OfourCpromptfermi\\ 
($\theta_j = 15^\circ$) & & & & & (\OfourCafterradiodeg) & (\OfourCafteropticdeg) & (\OfourCafterxdeg) & (\OfourCpromptswiftdeg) & (\OfourCpromptfermideg) \\ 
Rate DD2, $\chi_\mathrm{BH} = [0,0.5]$  & \OfourDLVKduty & \OfourDknJ & \OfourDknz & \OfourDkng & \OfourDafterradio & \OfourDafteroptic & \OfourDafterx & \OfourDpromptswift & \OfourDpromptfermi\\ 
($\theta_j = 15^\circ$) & & & & & (\OfourDafterradiodeg) & (\OfourDafteropticdeg) & (\OfourDafterxdeg) & (\OfourDpromptswiftdeg) & (\OfourDpromptfermideg) \\ 
\hline
\textbf{LVKI O5} & ~ & ~ & ~ & ~ & ~ &  &  &  & \\
Limit  & 12 & 21 & 24.4 & 26 & 0.01 & 26 & $10^{-13}$ & $3.5$  & $4$  \\ 
Rate SFHo, $\chi_\mathrm{BH} = 0$  & \OfiveALVKIduty & \OfiveAknJ & \OfiveAknz & \OfiveAkng & \OfiveAafterradio & \OfiveAafteroptic & \OfiveAafterx & \OfiveApromptswift & \OfiveApromptfermi \\ 
($\theta_j = 15^\circ$) & & & & & (\OfiveAafterradiodeg) & (\OfiveAafteropticdeg) & (\OfiveAafterxdeg) & (\OfiveApromptswiftdeg) & (\OfiveApromptfermideg) \\ 
Rate SFHo, $\chi_\mathrm{BH} = [0,0.5]$  & \OfiveBLVKduty & \OfiveBknJ & \OfiveBknz & \OfiveBkng & \OfiveBafterradio & \OfiveBafteroptic & \OfiveBafterx & \OfiveBpromptswift & \OfiveBpromptfermi \\ 
($\theta_j = 15^\circ$) & & & & & (\OfiveBafterradiodeg) & (\OfiveBafteropticdeg) & (\OfiveBafterxdeg) & (\OfiveBpromptswiftdeg) & (\OfiveBpromptfermideg) \\ 
Rate DD2, $\chi_\mathrm{BH} = 0$  & \OfiveCLVKduty & \OfiveCknJ & \OfiveCknz & \OfiveCkng & \OfiveCafterradio & \OfiveCafteroptic & \OfiveCafterx & \OfiveCpromptswift & \OfiveCpromptfermi\\ 
($\theta_j = 15^\circ$) & & & & & (\OfiveCafterradiodeg) & (\OfiveCafteropticdeg) & (\OfiveCafterxdeg) & (\OfiveCpromptswiftdeg) & (\OfiveCpromptfermideg) \\ 
Rate DD2, $\chi_\mathrm{BH} = [0,0.5]$  & \OfiveDLVKduty & \OfiveDknJ & \OfiveDknz & \OfiveDkng & \OfiveDafterradio & \OfiveDafteroptic & \OfiveDafterx & \OfiveDpromptswift & \OfiveDpromptfermi\\ 
($\theta_j = 15^\circ$) & & & & & (\OfiveDafterradiodeg) & (\OfiveDafteropticdeg) & (\OfiveDafterxdeg) & (\OfiveDpromptswiftdeg) & (\OfiveDpromptfermideg) \\ 

\end{tabular}
\label{tab:det_rates}
\tablefoot{The GW detection limits refer to the $\mathrm{SNR_{net}}$ threshold. Near infrared and optical limiting magnitudes are in the AB system; radio limiting flux densities are in mJy @ 1.4 GHz; X-ray limiting flux densities are in erg cm$^{-2}$ s$^{-1}$ keV$^{-1}$ @ 1 keV; gamma-ray limiting photon fluxes are in photons cm$^{-2}$ s$^{-1}$ in the 15--150 keV (\textit{Swift}/BAT) or 10--1000 keV (\textit{Fermi}/GBM) band. Detection rates are in $\mathrm{yr}^{-1}$. The reported errors, given at the 90\% credible level, stem from the uncertainty on the overall merger rate, while systematic errors are not included.}
\end{table*}

\section{Detailed study of the detectable EM signals}
\label{sec:detailed}

We now turn to detailed studies of the different EM counterparts.

\subsection{Kilonova}
\label{sec:kn}

\begin{figure*}
    \centering
    \includegraphics[width=\textwidth]{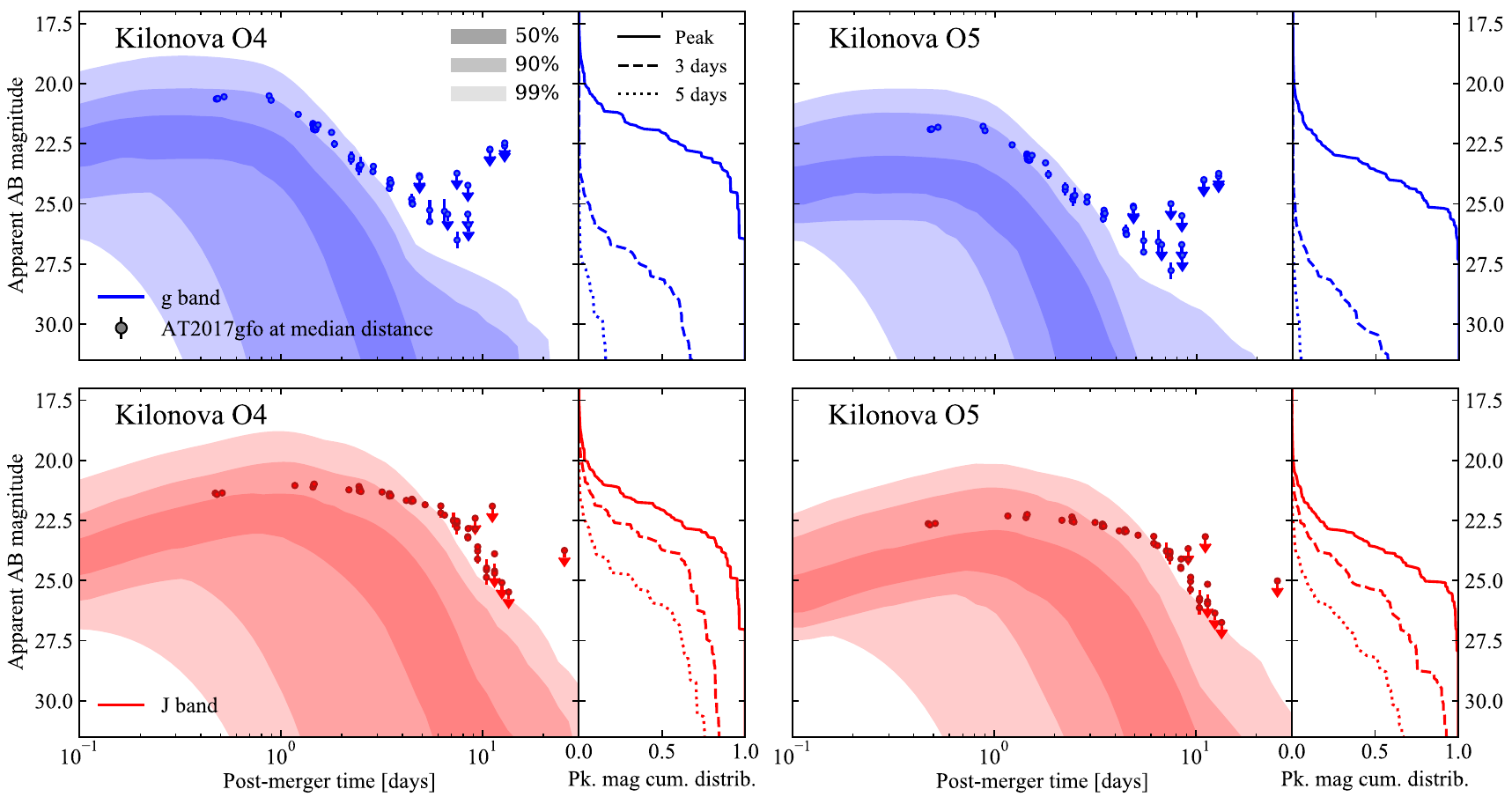}
    \caption{Distribution of O4 (left-hand panel) and O5 (right-hand panel) KN $g$ band (484 nm, upper panel in blue) and $J$ band (1250 nm, lower panel in red) magnitudes as a function of time. The shaded regions show the apparent AB magnitude versus post-merger time for $50\%$, $90\%$ and $99\%$ of our simulated KN light curves GW-detectable sources. Colored circles show extinction-corrected AT2017gfo data rescaled to the median distance of the GW triggers ($\sim$ 212 Mpc in O4, $\sim$ 381 Mpc in O5).
    The solid, dashed and dotted lines show the cumulative distributions of apparent magnitude at peak, at 3 days and at 5 days after the merger, respectively.}
    \label{fig:kn}
\end{figure*}

In Fig.~\ref{fig:kn}, we plot the distributions of the $g$ and $J$ band KN light curves of the GW-detectable events for the O4 and O5 runs, with the photometry of the KN signal associated with GW170817, AT2017gfo, at the median distance of the GW-detectable events with $m_\mathrm{out}>0$  ($\sim 212$ Mpc in O4, $\sim 381$ Mpc in O5). We took the photometric data from \citet{2017ApJ...851L..21V}. We note that, for the first few days, the typical BHNS KN light curve is at least two magnitudes dimmer than AT2017gfo, signaling the intrinsic weakness of the BHNS KN with respect to BNS KN sources at the origin of the low detection prospects described in Sec.~\ref{sec:mm}.

By applying the detection threshold we considered in O4 and O5 (22 and 26 magnitudes for the $g$ band, respectively and 21 for the $J$ band), we find that most of the KN\ae\ are undetectable in O4, whereas a majority of them are detectable in O5 in the $g$ band. Indeed, the magnitude threshold is slightly above the median peak magnitude of O4 and largely below for O5. This explains the jump in KN-detectable fraction that was found in Fig.~\ref{fig:detection_rates} and the lesser effect of the EoS and spin distributions in Fig.~\ref{fig:pop_variations}.

However, the distribution of magnitudes at 3 days post-merger shows that, even the deep limits considered in O5 are too shallow to detect the signal by this time. In fact, the distribution of light curves suggests that the post-peak dimming is even faster than for AT2017gfo. This poses the well-known issue of detecting the KN counterpart in time before it dims away, which has been a limiting factor in follow-up searches, notably because of the large GW sky maps. This issue is partially solved by considering infrared bands, where the signal is longer-lived. However, our numbers show that the weaker sensitivity of instruments, e.g. in the $z$ and $J$ bands, hinders the detectability prospects in these bands. This can be solved by large-FoV optical instruments such as the dedicated ZTF or survey instruments with a ToO program such as the Rubin Observatory \citep{2022ApJS..260...18A}. An additional solution to the recession of the GW horizon and the dimming KN\ae\ is to search for the non-thermal counterparts such as the radio afterglow directly with tiling instruments, a strategy that is available to upcoming radio surveys and that we study further in Sec.~\ref{sec:direct}.

\subsection{GRB afterglow}
\label{sec:after}

\begin{figure*}
    
    \centering
    \includegraphics[width=\textwidth]{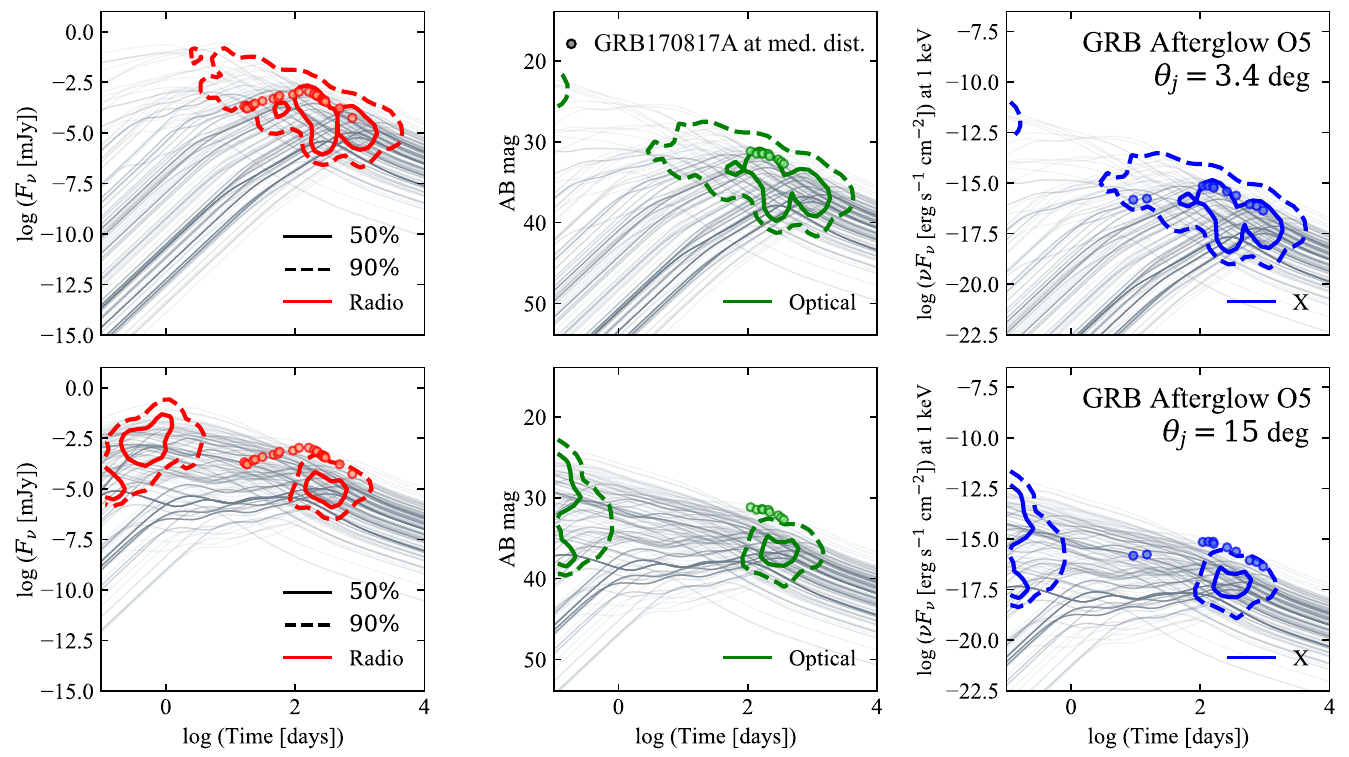}
    \caption{$F_\nu$, AB magnitude, and $\nu F_\nu$ versus time for the GRB afterglow light curves associated with O5-detectable sources in our population. In the top panel we assume a jet core half-opening angle $\theta_j = 3.4^{\circ}$: in the lower panel we assume $\theta_j = 15^{\circ}$. Solid and dashed contours contain $50\%$ and $90\%$ of the peaks, respectively. Red, green and blue colors indicate the radio ($1.4 \times 10^9$Hz), optical ($4.8 \times 10^{14}$Hz), X-ray ($2.4 \times 10^{17}$Hz) bands, respectively. The colored circles are the observed data of GRB170817A \citep{Makhathini2021} at the median distance of the GW triggers. The gray lines in the background are 300 randomly sampled light curves in the respective bands.}
    \label{fig:after}
\end{figure*}

In Figure \ref{fig:after} we show the properties of GRB afterglows associated with O5 GW-detectable binaries in our population by showing the contours containing $50\%$ (solid lines) and $90\%$ (dashed lines) of GRB afterglow peaks on the $F_{\nu}$ (AB magnitude, $\nu F_{\nu}$) versus observer time plane. The red, green, and blue colors refer to the radio, optical, and X-ray bands, respectively.  In the upper panel, we assume a jet half opening angle of $\theta_j = 3.4^\circ$ and in the lower panel of $\theta_j = 15^\circ$. 
In the narrow jet population, most peak times are at $\geq10$ days, with a small subsample peaking at early times ($\sim$ hours) in the optical and X-rays, producing very bright emission, due to on-axis events with small viewing angles. In the population with a larger opening angle of $15^\circ$, more jets can fall within the viewing angle of the observer. As a result, the number of on-axis events will increase compared to the narrow jet population, leading to the bimodality of the distribution of afterglow peaks at short and long times apparent in the bottom panels of Fig.~\ref{fig:after}.
In order to help visualize the underlying light-curve behavior, we display 300 randomly sampled afterglow light curves from our GW triggers (thin gray lines) in the background (the band is the same as the respective contours). The small circles indicate GRB 170817A data \citep{Makhathini2021} at the median distance of our population of GW triggers with $m_\mathrm{out}>0$ ($\sim$ 381 Mpc in O5), whose peak lies within the $50\%$ ($90\%$) contours in all three bands, assuming $\theta_j = 3.4^\circ$ ($\theta_j = 15^\circ$). 

\subsection{GRB prompt}
\label{sec:grb}

\begin{figure}[t]
    \centering
    \includegraphics[width=\columnwidth]{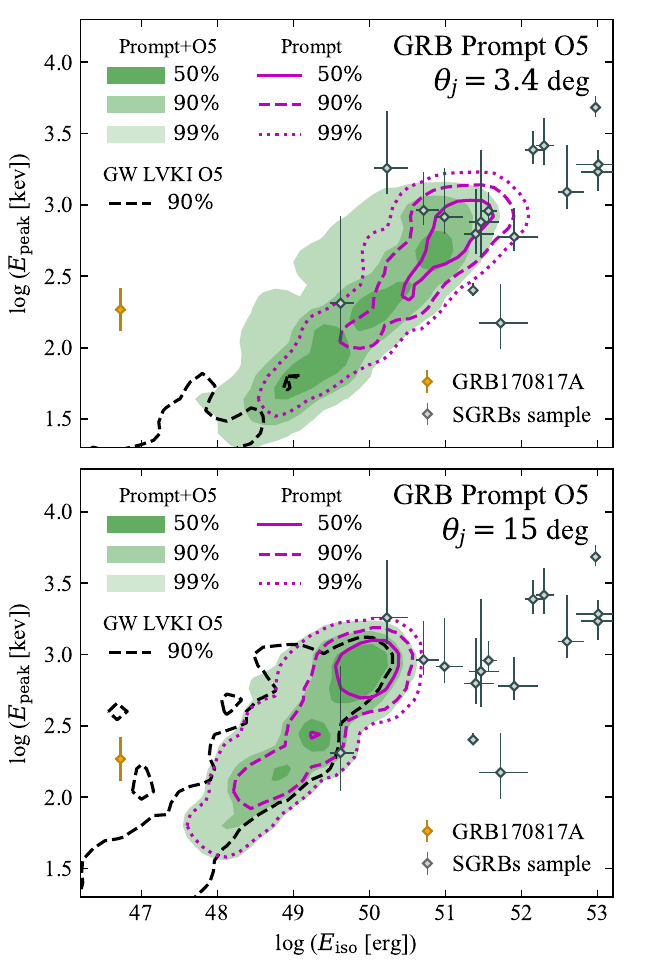}
    \caption{Rest-frame SED peak photon energy $E_\mathrm{peak}$ versus the isotropic-equivalent energy $E_\mathrm{iso}$ for our BHNS population. In the top panel we assume a jet core half-opening angle $\theta_j = 3.4^{\circ}$, in the lower panel we assume $\theta_j = 15^{\circ}$. The filled green colored regions contain $50\%$, $90\%$ and $99\%$ of the binaries both GRB Prompt- and LVKI 05-detectable. The magenta lines contain $50\%$, $90\%$ and $99\%$ (solid, dashed and dotted, respectively) of the GRB prompt-detectable binaries. The black dashed line contains $90\%$ of the O5-detectable binaries. The black dots with error bars represent a SGRB sample for comparison \citep{Salafia2023}. The orange dot is GRB170817A.}
    \label{fig:prompt}
\end{figure}

In order to visualize the GRB prompt emission parameter space accessible by multi-messenger observations, in Figure \ref{fig:prompt} we show how the rest-frame spectral energy distribution (SED) peak energy $E_\mathrm{peak}$ and the isotropic-equivalent energy $E_\mathrm{iso}$ are distributed for events that satisfy our detectability criteria for the GW signal, the GRB prompt emission, or both. The two variations in the jet half-opening angle are displayed in two separate panels: $\theta_j = 3.4^\circ$ (upper panel) and $\theta_j = 15^\circ$ (lower panel). Green filled contours refer to events that can be jointly detected by the O5 LVKI network and \textit{Fermi}/GBM: different shades of green contain a progressively higher fraction ($50\%$, $90\%$, and $99\%$) of the joint GRB prompt- and O5-detectable binaries. The black dashed contour contains 90\% of the events that are detectable in GW. The magenta contours contain 50\%, 90\% and 99\% of events detectable by \textit{Fermi}/GBM.  

For comparison with the known cosmological population, we show with gray diamonds the properties of a sample of short GRBs (SGRBs) with known redshift \citep{Salafia2023}. The orange diamond in the plot corresponds to the position of GRB 170817A.

Given the monotonic dependence on the viewing angle of both $E_\mathrm{iso}$ and $E_\mathrm{peak}$, as shown in more detail in Appendix \ref{sec:appB}, the GRBs in our model naturally feature an "Amati" correlation \citep{Amati2002,Tsutsui2013}, with events on the upper-right part of the plane observed close to on-axis and events with lower $E_\mathrm{iso}$ and $E_\mathrm{peak}$ observed farther from the jet axis \citep[see][for a detailed explanation of why and how this kind of correlation is induced by the presence of an angular jet structure]{Salafia2023}. The contours show clearly that Fermi/GBM preferentially detects events close to on-axis (actually, close to $\theta_\mathrm{j}$ where the accessible solid angle is maximised); the GW detectors preferentially detect events with a larger viewing angle (peaking at around $30^\circ$ due to a trade-off between the larger solid angle and the weaker GW strain with increasing viewing angle, e.g.\ \citealt{Schutz2011}); joint GRB plus GW detections occupy a region in between these two, especially for $\theta_\mathrm{j}=3.4^\circ$ where the peaks of the GW and GRB  distributions are more widely separated.

The main impact of the two different jet core half-opening angle assumptions on these distributions consists in a horizontal shift towards lower $E_\mathrm{iso}$ for increasing $\theta_\mathrm{j}$. This is a consequence of the fact that we keep the total jet energy fixed when varying the opening angle, which implies that the maximum attainable $E_\mathrm{iso}$ for each jet (which corresponds to that measured by an on-axis observer) scales as $\theta_\mathrm{j}^{-2}$. This shows that, if the average opening angle of BHNS jets were indeed larger than that of BNS jets, and if the spectral properties of the prompt emission were otherwise similar as we assumed, then BHNS-associated GRBs would follow a distinct Amati correlation with respect to BNS-associated ones. This statement clearly rests on very uncertain assumptions and must therefore be taken with a grain of salt.

Inspection of Table \ref{tab:det_rates} additionally shows that a larger $\theta_\mathrm{j}$ positively impacts the joint GRB plus GW detection rate, despite the GRBs being dimmer overall. This is a consequence of the fact that, for the current generation of GW detectors, the horizon for a joint detection is set by the GW than by the GRB: while the latter can be detected in principle out to $z\gtrsim 2$, the former are currently accessible only out to $z\sim 0.1-0.3$ (see Figure \ref{fig:detection_rates}).

\section{Discussion}
\label{sec:discussion}

\subsection{Comparison with other works with similar goals}

As pointed out by \citet{Boersma2022}, the distribution of BH spins plays a crucial role in determining the prospects of  MM detections of BHNS mergers.
This role is more important than the NS EoS, and the uncertainty due to the unknown spin distribution surpasses the uncertainty on the intrinsic BHNS merger rate. However, in contrast to \citet{Boersma2022}, we find that the prospects for radio afterglow detections in the upcoming runs of the Ligo-Virgo-generation of GW interferometers are not that poor, assuming the radio instruments follow their planned development. In the O5 run with a limiting radio sensitivity of $0.01~{\rm mJy}$, the detection rate  is no lower than 1 every few years in the fiducial model where BHs are non spinning, and can reach order $1\,{\rm yr}^{-1}$ in  the case BHs are endowed with a moderate spin (see Tab.~\ref{tab:det_rates}). The issue of localizing the source in order to discover the afterglow remains since we find that the likeliness of detecting a short GRB counterpart is low, as in \citet{Boersma2022}, and more so in O5 than in O4. Even with the current generation of IFOs, the coming online of more GW detectors will decrease the expected size of the GW sky maps, and the deployment of more optimized follow-up strategies can help meet the challenge of covering them.

In their KN modeling,  \citet{2021ApJ...917...24Z} follow an approach similar to ours when informing the parameters of a semi-analytical model (e.g., mass and opacity of various ejecta components) with numerical simulations. For the O5 run, we find a similar limiting magnitude of 23--24 in the visible bands found by \citet{2021ApJ...917...24Z} to recover half of the KN counterparts, with the spinning-BH hypothesis (see Tab.~\ref{tab:det_rates}). We also share their conclusion of the fast decay of these signals, possibly hindering their detection even three days post-merger. A possible way to circumvent this issue is to search for the afterglow using wide-FoV instruments to tile the GW sky map, as suggested in the X-rays by \citet{2022A&A...665A..97R} and as we discuss in Sec.~\ref{sec:direct}.

Conversely, \citet{Zhu2022} find a much larger rate of combined GW, GRB prompt and KN detections in O4 and O5 than we do. Interestingly, their study follows a very different method, which relies on the parameter inference of three long-duration GRBs that are hypothesized to be of BHNS-merger origin. Their analysis suggests massive-BH and highly spinning BH progenitors, which could explain their much higher detection rates. Of course, this conclusion is highly dependent on the emission model they use to infer the properties of the progenitors of these three bursts. In any case, the comparison of our formation-channel-based work with studies such as \citet{Zhu2022} underlines the importance of considering BHNS systems in all of their manifestations: as GW triggers with EM counterparts but also as the progenitors of a fraction of SGRBs (\citealt{2020ApJ...895...58G}, or even some LGRBs, \citep[recently, e.g.,][]{2022Natur.612..223R,2023NatAs...7..976L}), for which the longer observation history can be an advantage over the GW triggers as a tool to study the population of BHNS mergers.

\subsection{Direct searches for the radio afterglow with a titling strategy}
\label{sec:direct}

\begin{table*}
\begin{center}
\caption{Relative number of GW-detectable systems with detectable GRB afterglows (radio band), with and without detectable KN, for our fiducial population model, in an O5-type run, and the two jet opening angle hypotheses.}
\label{tab:ag}
\begin{tabular}{lcc}
\hline \hline
Fractions of GW triggers with detectable radio afterglow,  & $\theta_j = 3.4^{\circ}$ & $\theta_j = 15^{\circ}$\\
   among which:      & O5 &  O5 \\ \hline
KN detectable & 67\% &  74 \%\\
KN undetectable & 33\% & 26\% \\
KN undetectable \& GRB prompt undetectable  &  12 \%  &  15\% \\ 
\end{tabular}\\
\end{center}
\tablefoot{The threshold for GRB radio afterglow detection considered was $0.01\,{\rm mJy}$. KN detection is defined by detection in $g$ (24) or in $z$ (22.5), see text for more details.}
\end{table*}

In Sec.~\ref{sec:mm}, we studied the detectability of the radio afterglow  of merging BHNS systems at a fixed detection threshold of the radio flux. In practice, this criterion assumes knowing the location of the source, e.g., thanks to a detection of the KN before the afterglow searches. In the O2 and O3 GW observing runs, searching for the KN signal was seen as a stepping-stone to locate the event before performing deep searches for other EM counterparts, such as the relativistic jet's afterglow. Efforts in developing large-FoV optical instruments and networks of instruments were partly driven by the wish to ensure this KN detection, overcoming the potentially large uncertainties of GW sky maps. 

However, as shown in the O3 run, and for those events likely hosting a NS, covering the GW sky maps in optical searches is challenging (we refer to GW190425 and the two BHNS systems; see Sec.~\ref{sec:intro} and, e.g., \citealt{Coughlin2019,Abbott2021d}). In addition, our results show that most KN\ae\ associated with BHNS systems remain undetectable for the majority of the GW triggers (see Tab.~\ref{tab:det_rates} and Fig.~\ref{fig:kn}). This should also be the case for BNS systems \citep[e.g.][]{Colombo2022}. Furthermore, detecting a KN is challenging considering its fast decay. This limitation motivates the development of new follow-up strategies to detect the afterglow emission without having first acquired the source's location thanks to the KN detection.

Luckily, the long duration of the radio afterglows due to the system's off-axis inclination angles, as exemplified in Fig.~\ref{fig:after}, represents a significant advantage over the shorter-lived KN\ae\, suggesting the  direct search of the afterglow counterpart after a GW trigger with optimized follow-up strategies.

To quantify the gain in discovering the afterglow in absence of a KN counterpart we show in Tab.~\ref{tab:ag} the number of detectable BHNS afterglows with and without a detectable KN in an O5-type run. We considered a radio afterglow threshold of $0.01\,{\rm mJy}$ for O5 as in Sec.~\ref{sec:mm} and KN limits of 24 in $g$ and 22.5 in $z$. In Sec.~\ref{sec:mm}, we had considered a $g$-magnitude limit of 26 respectively for O5, representing the performance of a single deep instrument. However, due to limitations in sky coverage and instrument availability, it is likely that networks of optical instruments with shallower imagery will continue to play a significant role in kilonova discovery searches. Therefore, a more realistic estimate of the overall follow-up performance in O5 is a shallower limit of 24 in $g$.

The table shows that, in O5, 33\% of GW triggers with a detectable afterglow do not present a detectable KN for $\theta_j = 3.4^\circ$, and 26\% for $\theta_j = 15^\circ$. This suggests that relying on the KN as a stepping-stone for afterglow searches limits the potential for afterglow detection. As expected, inspection of this subclass of events shows that the afterglows without KN are mostly high-distance events with low inclination angles, close to the GW horizon. In addition, 12--15\% of these afterglows present neither a KN nor a detectable GRB prompt emission, prohibiting the GRB prompt from providing localization data.

\begin{figure}
\includegraphics[width=\linewidth]{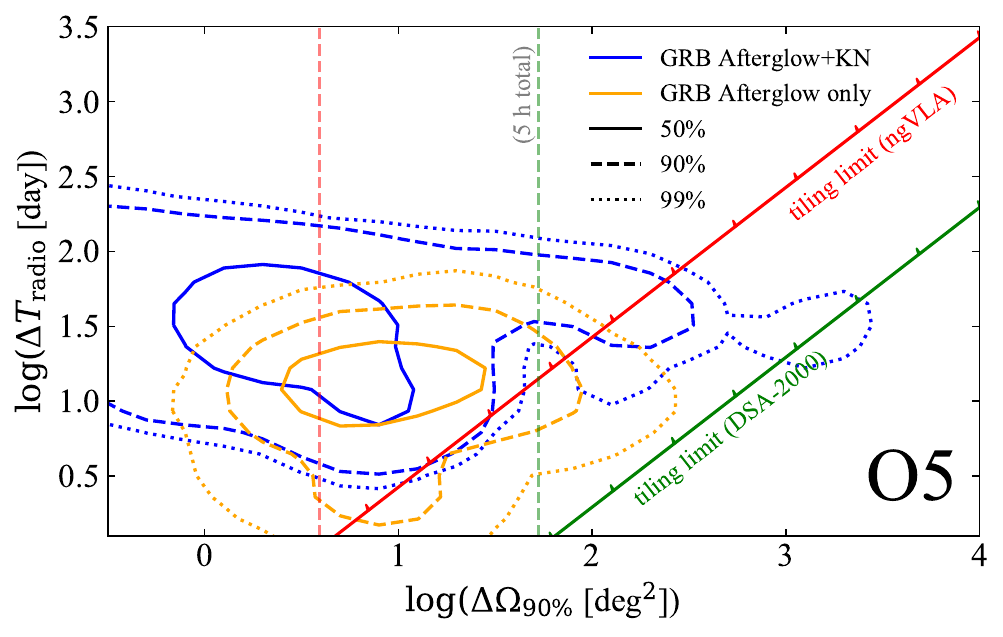}
\caption{Population contours of the GW-detectable events in O5 with detectable radio afterglow in the plane of detectability interval of the afterglow versus 90\%-confidence GW sky map size. We distinguish the events with a detectable (blue) or undetectable (orange) KN. The limiting fluxes and magnitudes for the radio afterglows and KN are as in Tab.~\ref{tab:ag}. The green and red lines indicate the limit above which the tiling strategy is viable for the DSA-2000 and the ngVLA, respectively.  That is, the GW sky map can be covered at least 5 times by tiles over the afterglow time of observability (Eq.~\ref{eq:tiling}). The vertical lines indicate the GW sky map sizes that require a total observing time of one and five hours to cover by the two instruments, using the same color code.} 
\label{fig:ag}
\end{figure}

To bypass this dependence on the KN signal, we explore the possibility that large-FoV radio instruments can tile the GW sky map in search for the afterglow, as already considered for future radio arrays by \citet{2021MNRAS.505.2647D} and  in the X-ray band for upcoming large-FoV instruments by \citet{2022A&A...665A..97R}. In Fig.~\ref{fig:ag}, we show the contours of GW triggers with detectable radio afterglows in O5 runs, by distinguishing those with and without detectable KN, using the same radio and KN limits as in Tab.~\ref{tab:ag}. The orange contours corresponds to the KN-less afterglows discussed in Tab.~\ref{tab:ag}, which could be the target of this radio tiling strategy.

In this plot, a tiled search for the afterglow is only viable under the following condition:
\begin{equation}
\frac{\Delta T_{\rm radio}}{\Delta t_{\rm tile}} > N \times \frac{\Delta \Omega_{\rm GW}}{\Delta \Omega_{\rm FoV}}
\label{eq:tiling}
\end{equation}
such that the GW sky map of size $\Delta \Omega_{\rm GW}$ can be covered entirely at least $N$ times by tiles of size $\Delta \Omega_{\rm FoV}$ in the time that the afterglow is detectable. Being able to cover the GW sky map more than once ensures enough photometry points to identify the afterglow among various transient sources can be acquired. We added this integer $N$, which increases the observation time required to tile a GW sky map, also to capture the unmodelled effects of overheads in observations and gaps in instrument availability; one can consider a fiducial value of $N = 5$. Here, $\Delta t_{\rm tile}$ is the tiling period, that is, the time interval between two tiles, which we take to be the integration time for radio imagery, neglecting repointing time and similar overheads. $\Delta T_{\rm radio}$ is the afterglow detectability time interval, i.e., the time length during which the afterglow flux is above the radio threshold of the given instrument, as determined from our afterglow model (Sec.~\ref{sec:EMmodel}).

This tiling strategy is especially interesting for large-FoV instruments. We follow \citet{2021MNRAS.505.2647D} in considering the DSA-2000 \citep{2019BAAS...51g.255H} and the Next-Generation VLA (ngVLA, \citealt{2018SPIE10700E..1OS}), both instrument concepts with plans for a target-of-opportunity program. The DSA-2000 is a survey instrument with a single-image FoV of $\Delta \Omega_{\rm FoV} =  10.6\,{\rm deg}^2$ that can reach an rms sensitivity of $1\,\mu{\rm Jy}$ in $\Delta t_{\rm tile} = 1\,{\rm h}$ in the 2\,GHz band \citep{2019BAAS...51g.255H}. The ngVLA is a general-purpose radio array with a smaller FoV of $\Delta \Omega_{\rm FoV} =  0.13\,{\rm deg}^2$ and a shorter time of $\Delta t_{\rm tile} = 10\,{\rm min}$ to reach $1\,\mu{\rm Jy}$\footnote{The announced performance for the ngVLA is $0.41\,\mu{\rm Jy}$ in one hour \citep[][Tab.~1]{2018SPIE10700E..1OS}, equivalent to $1\,\mu{\rm Jy}$ in 10 min, considering $F_{\rm lim} \propto t^{-1/2}$.} in the same band.

In the radio tiling strategy, the critical factor is the tiling speed down to a given sensitivity:
\begin{equation}
\Sigma = \frac{\Delta \Omega_{\rm FoV}}{\Delta t_{\rm tile}},
\end{equation}
which, is $\Sigma_{\rm DSA} = 10.6\,{\rm deg}^2/{\rm h}$ and $\Sigma_{\rm ngVLA} = 0.78\,{\rm deg}^2/{\rm h}$ for the two instruments, respectively, down to $1\,\mu{\rm Jy}$. 

In Fig.~\ref{fig:ag}, the green and red lines show the tiling limit according to Eq.~\ref{eq:tiling} for the DSA-2000 and ngVLA. The dashed vertical lines indicate the GW sky map sizes that can be covered in five hours of total tiling observation by these instruments. We find that all of the KN-less GRB afterglows are in reach of the tiling strategy of DSA-2000 and most of them are in reach for ngVLA, thanks to their large tiling speeds. The limits of 1\,$\mu{\rm Jy}$ for DSA-2000 and ngVLA, an order of magnitude deeper than the detection limits applied in Fig.~\ref{fig:ag}, would allow these instruments to make a good sampling of the light curves and identify the GRB afterglow transient. We note that, similarly to KN searches, once a transient candidate is identified by the radio searches, these candidates can be announced and monitored by other instruments, sharing the load of follow-up observations. In this scenario, the monitoring of radio candidates identified early enough can even lead to the KN detection through optical searches.

In O5, we find that these KN-less GRB afterglows will be detectable for up to tens of days, leaving more than enough time to tile the sky maps. The prominence of the KN-less  afterglows at larger sky map sizes over the afterglows with KN further motivates this strategy. We note that, even in O4 where the KN\ae\ are in principle detectable, this strategy opens an alternate channel for counterpart discovery in the case that the optical network did not detect the detectable KN beforehand, e.g., for lack of instrument availability or ToO program time. In O5, most of the GRB radio afterglows are discoverable with a total observing time of less than 5 hours for DSA-2000, corresponding to 5 tiles of DSA-2000. A significant fraction of events are discoverable within 5 hours by ngVLA, corresponding to 30 tiles for this instrument.

The potential for radio tiling to improve the outcome of EM searches is critical for systems such as BHNSs for which the detection prospects are tight (see Sec.~\ref{sec:mm}). The DSA-2000 and ngVLA concepts project to carry out target-of-opportunity programs. This ToO time could be used for the GRB afterglow discovery of already-localized events, as was the case for radio arrays for GW170817. Alternatively, this time could be condensed into the tiling of GW sky maps. With this second option, a radio array can allow the discovery of up to 100\% more GRB afterglow counterparts than if the community relies on the detection of the KN first.

In any case, like all tiling strategies, the radio tiling strategy requires archival images of the sky to flux limits as deep as the follow-up searches. This is required to carry out difference-imaging and identify the transient sources. To our knowledge, such archival images do not exist in the GHz band. However, the instruments that are best fit for tiling strategies are survey instruments, such as the upcoming DSA-2000, which could then rely on its own survey images for the searches in its ToO program.

Whether targeting the KN or the GRB afterglow, the choice of events to follow up and when to start tiling could benefit from low-latency release of GW constraints on the binary parameters. In particular, the early publication of the binary masses would help  determine the expected afterglow flux, and information on the inclination angle would help predict the peak time or optimal interval for observation. For the KN, follow-up teams have already developed models to generate the light curved likely to arise from a source, given the binary parameters \citep{Barbieri2020,2021MNRAS.505.3016N}. A similar tool for afterglows could help optimize searches, if the GW constraints on the binary systems are released in low latency.

\section{Summary and conclusions}
\label{sec:conclusion}

In this article, we studied the MM detectability of merging BHNS binaries in the O4 and O5 runs of the global GW detector network.  BHNS systems are MM sources and here we explored their observability by characterizing their multi-frequency EM emission and computing the cumulative MM detection rates. We started by constructing a synthetic population based on the binary evolution models from \citet{Broekgaarden21}, assuming two different BH spin prescriptions and two variations of the NS EoS. We computed the expected properties of the material expelled during and after the coalescence using fitting formulae calibrated on a large set of relativistic hydrodynamical simulations. We then used this information as input to semi-analytical models to compute the observable properties of the associated KN, multi-wavelength GRB afterglow and GRB prompt emission. This allowed us to construct the distributions of the EM observables for O4 and O5 GW-detectable events.

Overall, the prospects for O4 are poor, with less than one detectable KN counterpart every few years when assuming non-spinning BHs in the binaries. Under the assumption that  BHs can have spins up to $\chi_\mathrm{BH} = 0.5$, the prospects are of the order of one detectable counterpart per year, with the effect of the spin distribution on the rate of detectable counterparts being larger than that of the EoS.

As of O5, the prospects for MM observations will be brighter, thanks to the improved performance of the GW and EM instruments. We expect one detectable KN counterpart per year assuming non-spinning BHs, and about ten per year if the BHs are spinning. The prospects for detectable radio GRB afterglows follow the same trend, with one detectable counterpart assuming spinning BHs in O5.

The strong effect of the spin distribution on the EM counterparts opens the way to constraining the formation channels of the BHNS systems since these channels determine the distribution of binary parameters. However, it is only as of O5, when the potential detection rates will be significant, that the constraining power of MM observations will appear.

In any case, even when the KN\ae\ are detectable, their observation should remain challenging due to their fast dimming, and the large GW sky map sizes. This motivates tailored observing strategies aiming to provide an alternative channel to discover afterglow counterparts directly in the radio band, leveraging the afterglow's long duration, in case the KN was not detected. Large field-of-view radio arrays, such as the projected DSA-2000 and ngVLA, can effectively carry out a tiling of the GW sky map during their target-of-opportunity program. This radio tiling strategy could allow allow recovery of up to twice as more GRB afterglow counterparts of BHNS systems than if one relies on first detecting the KN.

\section*{Acknowledgments}

This work was initially thought out and developed at The Unconventional Thinking Tank Conference 2022, which is supported by INAF. AC thanks Matteo Barsuglia, Samuele Ronchini and Sylvia Biscoveanu for helpful discussions and comments on this manuscript. RD is supported by the European Research Council Advanced Grant "JETSET: Launching, propagation and emission of relativistic jets from binary mergers and across mass scales" (grant no. 884631). The work of MM is supported by European Union's H2020 ERC Starting Grant No.~945155--GWmining, Cariplo Foundation Grant No.~2021-0555, the ICSC National Research Centre funded by NextGenerationEU, and MIUR PRIN Grant No. 2022-Z9X4XS. The work of FI is supported by the  Swiss National Science Foundation, grant 200020$\_$191957,  by the SwissMap National Center for Competence in Research, and by the Istituto Svizzero "Milano Calling" fellowship.  Part of the computations made use of the Baobab and Yggdrasil clusters at the University of Geneva. FSB acknowledges support for this work through the NASA FINESST scholarship 80NSSC22K1601 and from the Simons Foundation as part of the Simons Foundation Society of Fellows under award number 1141468. TF acknowledges support for this work through the Swiss National Science Foundation (project numbers PP00P2\_211006 and CRSII5\_213497). \texttt{GWFAST} is publicly available at \url{https://github.com/CosmoStatGW/gwfast}.

\bibliographystyle{aa}
\bibliography{main}

\begin{appendix} 
\section{Mass distribution comparison} \label{sec:mass_comparison}

\begin{figure*}
\resizebox{\hsize}{!}{\includegraphics[width=\textwidth]{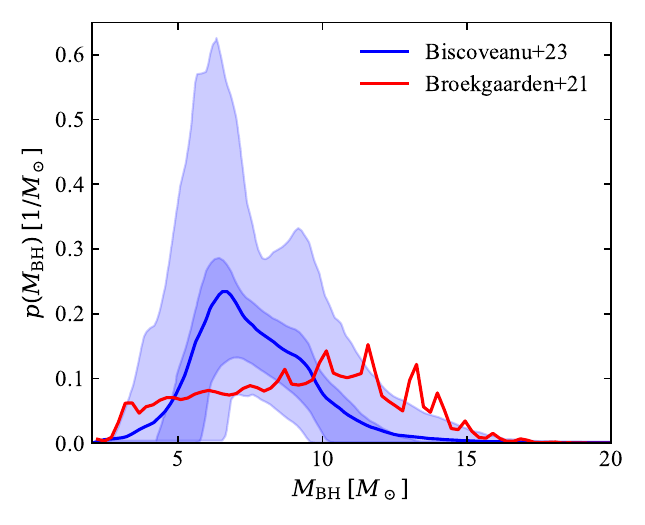}\includegraphics[width=\textwidth]{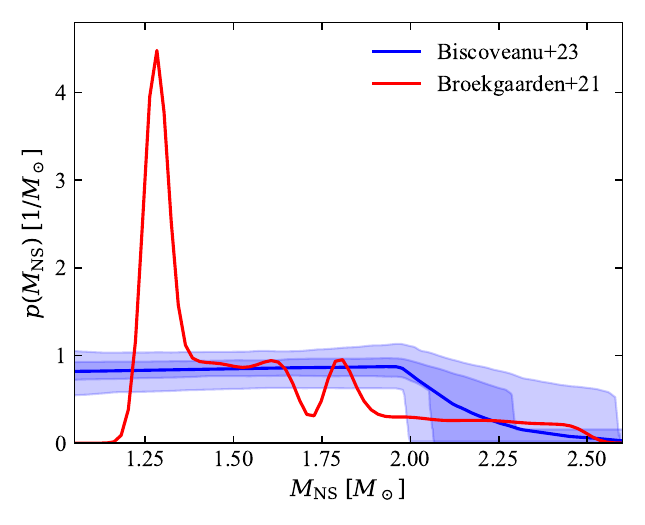}}
\caption{Component mass probability distribution comparisons. Component masses (left panel: BH; right panel: NS) probability distribution (red lines) for the fiducial model in \cite{Broekgaarden21} used in this work and posterior predictive distributions (solid blue), 50\% and 90\% credible intervals (shaded blue) inferred from the population of four candidate BHNS events detected in GW with a FAR rate $\leq 1$ yr$^{-1}$ \citep{Biscoveanu2023}.} 
\label{fig:mass_comparison}
\end{figure*}

It is informative to compare the mass probability distribution of our population model of BHNS systems with the one inferred from the GW data. Therefore, we show in Fig. \ref{fig:mass_comparison} a comparison of the probability distribution of primary (left panel) and secondary (right panel) component masses based on the population synthesis fiducial model from \citet{Broekgaarden21} assumed in this work (red lines) with the corresponding distributions obtained by \citet{Biscoveanu2023} (blue lines) using an approach driven by the population of four candidate BHNS events detected in GW so far with a FAR rate $\leq 1$ yr$^{-1}$. In particular the solid blue line represent the posterior predictive distribution and the shaded blue regions the 50\% and 90\% credible intervals, under a Gaussian mass ratio model. 

In the BH mass range of interest in this work $M_\mathrm{BH}\lesssim 11 M_\odot$, corresponding to systems that can power EM counterparts in our optimistic scenario (see Sec. \ref{sec:channels}), the mass distribution from \citet{Broekgaarden21} lies within the 90\% credible interval inferred from the GW data. Moreover, the requirement of having systems with $M_\mathrm{BH} \lesssim 5 M_\odot$ in our fiducial scenario is not in contrast with the distribution predicted by \citet{Biscoveanu2023}. 
The NS mass distribution inferred from the GW data is flatter than the one used in this work and they both fall off sharply at the maximum mass. The two peaks in the red curve are caused by the choice of supernovae remnant mass prescription in \cite{Broekgaarden21} and it is common to all the variations performed in the work.

Overall, we can conclude that the mass distributions of our BHNS population model are consistent with the first constraints deduced from GW data, lending credit to the basis of our MM study.

\section{Prompt emission properties as a function of the viewing angle}
\label{sec:appB}

\begin{figure*}
\resizebox{\hsize}{!}{\includegraphics[width=\textwidth]{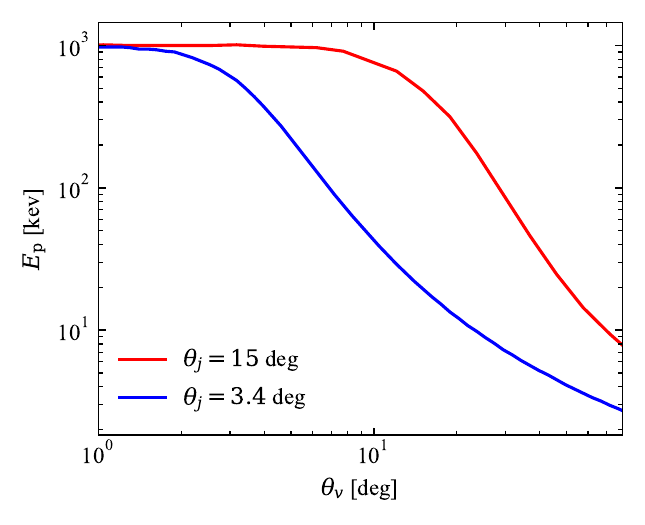}\includegraphics[width=\textwidth]{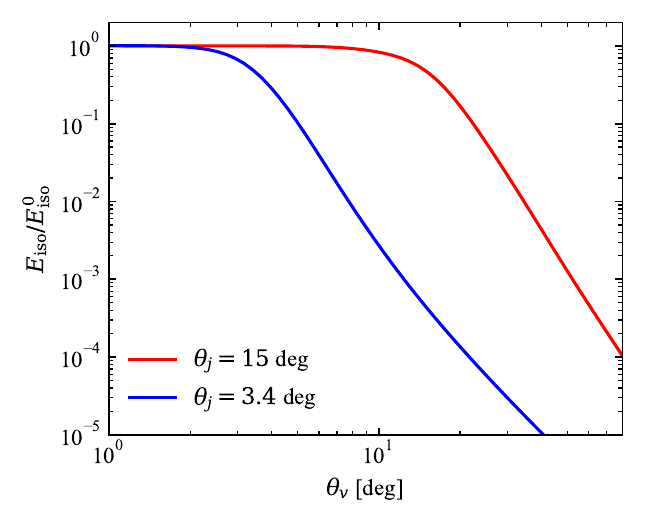}}
\caption{Rest-frame SED peak photon energy $E_\mathrm{peak}$ (left panel) and isotropic-equivalent energy $E_\mathrm{iso}$  normalized to the value $E_\mathrm{iso}^0$ measured by an on-axis observer (right panel), as functions of the viewing angle $\theta_v$. Blue and red colors refer to $\theta_j = 3.4^\circ$ and $\theta_j = 15^\circ$, respectively.} 
\label{fig:jet_structure}
\end{figure*}


In the left panel of Fig.\ \ref{fig:jet_structure} we show the rest-frame SED peak photon energy $E_\mathrm{peak}$ as a function of the viewing angle $\theta_v$. In the right panel we represent the isotropic-equivalent energy $E_\mathrm{iso}$ normalized to the value measured by an on-axis observer. Both are computed as detailed in Appendix B of \citet{Colombo2022}. 
The blue and red colors in Fig.\ \ref{fig:jet_structure} refer to the assumption on the jet half-opening angle: $\theta_j = 3.4^\circ$ and $\theta_j = 15^\circ$, respectively.

These plots can help understand the shifting of contours in Fig.\ \ref{fig:prompt}, depending on different assumptions on the jet half-opening angle. For a larger $\theta_j$, the detectable GRB prompt events concentrate at $E_\mathrm{peak} \sim 10^3$ keV and $E_\mathrm{iso} \sim 10^{50}$ erg. Instead the distribution for the smaller $\theta_j$ is much broader, with the same maximum $E_\mathrm{peak}$, but a maximum $E_\mathrm{iso} \sim 10^{51}$ erg. Looking at Fig. \ref{fig:jet_structure}, it's easy to understand why; for $\theta_j = 15^\circ$ more viewing angles are intercepted, leading to the maximum value both for $E_\mathrm{peak}$ and $E_\mathrm{iso}$. The reduction in the values of the isotropic-equivalent energy is due to the fact that $E_\mathrm{iso}$ is scaled based on the core energy $E_\mathrm{c} \propto \theta_j^{-2}$, while preserving the jet energy $E_\mathrm{jet}$.

\section{Sky localization and detection limits}
\label{sec:appC}

\begin{figure*}
    \centering
    \includegraphics[width=0.92\textwidth]{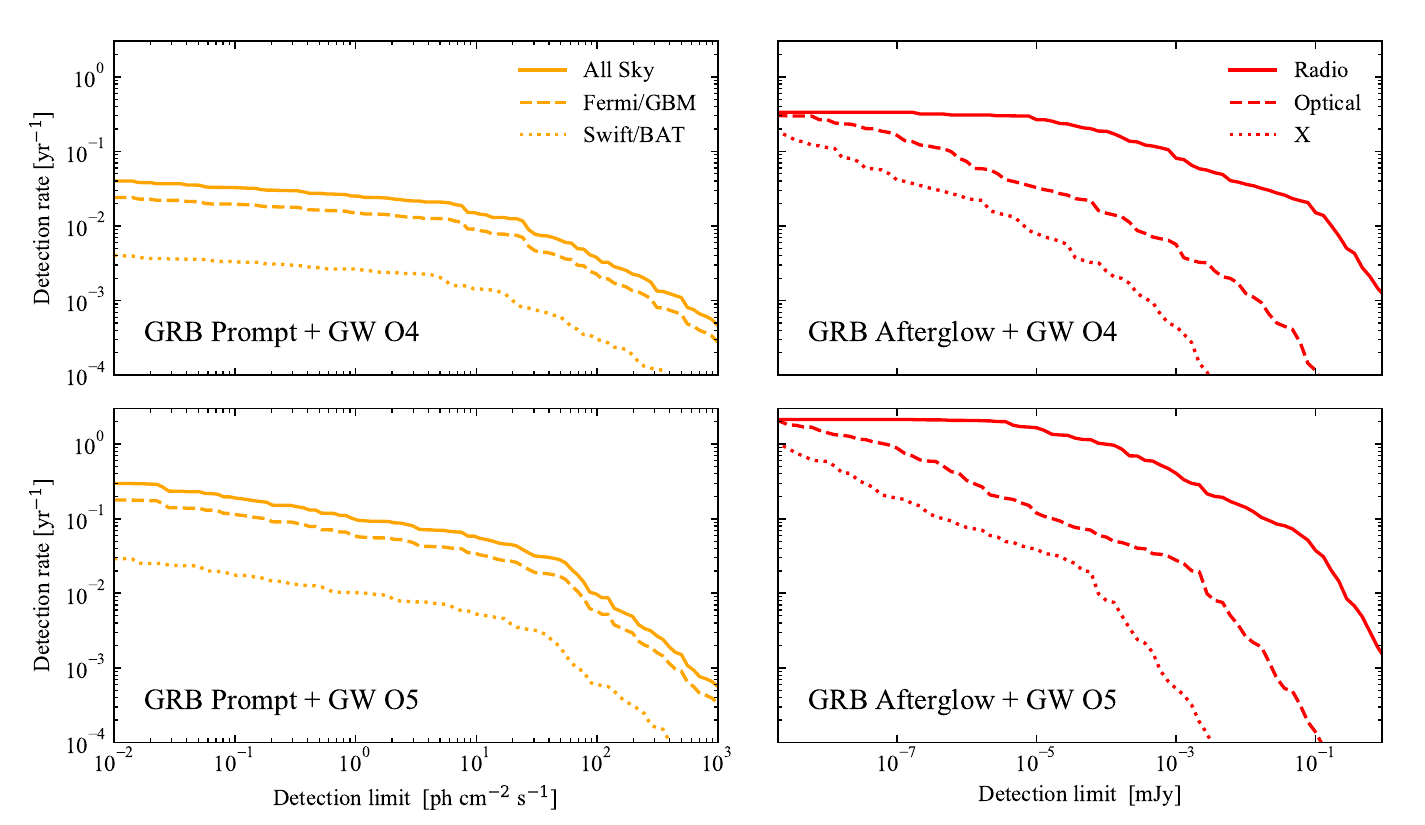}
    \caption{Detection rate as a function of detection limit threshold for our fiducial BHNS population. In the top panel we assume the LVK O4 detectors network, while in the lower panel we assume the LVKI O5 detectors, both with a 70\% duty cycle for each detector. The orange and red colors indicate respectively the GRB prompt+GW and the GRB afterglow+GW detectable binaries. In the left panel, the solid line indicates an all-sky field of view with a 100\% duty cycle, the dashed and dotted lines account for the \textit{Fermi}/GBM and Swift/BAT duty cycle and field of view, respectively. In the right panel, the solid, dashed, and dotted lines indicate the radio, optical, and X bands, respectively.}
    \label{fig:rate_limit}
\end{figure*}

\begin{figure*}
    \centering
    \includegraphics[width=0.92\textwidth]{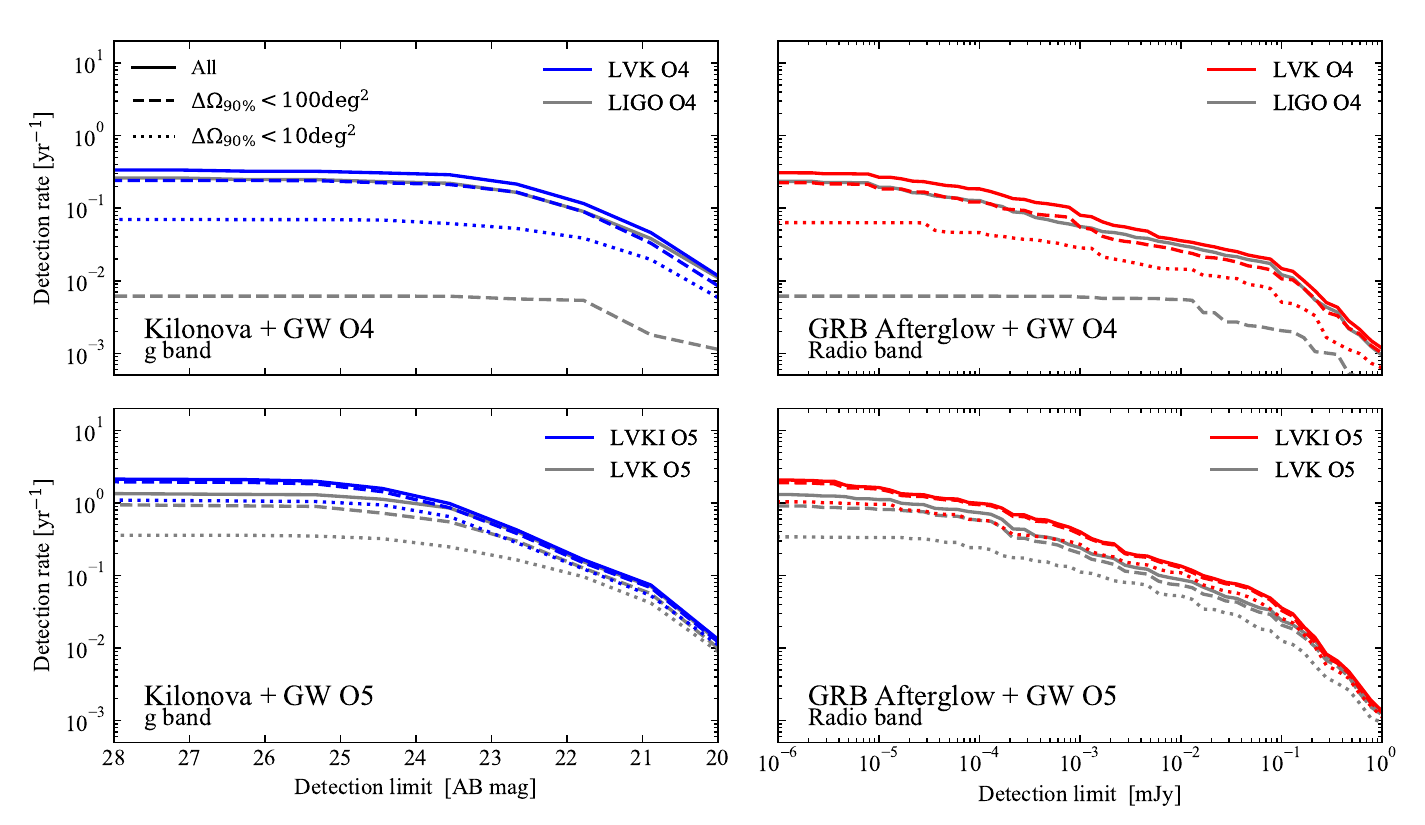}
    \caption{Detection rate as a function of detection limit threshold for our fiducial BHNS population. In the top panel we assume the LVK O4 detectors network (colored lines) and a network consisting of only two aLIGO (gray lines). In the lower panel we assume the LVKI O5 detectors network (colored lines) and the same network without LIGO--India (black lines). Every network is assumed with a 70\% duty cycle for each detector. The blue and red colors indicate respectively the kilonova+GW ($g$ band) and the GRB afterglow+GW (radio band) detectable binaries. The solid line indicates all the detectable binaries, the dashed and dotted lines the detectable binaries with $\Delta\Omega_{{\rm 90}\%}<100\mathrm{deg}^2$ and the ones with $\Delta\Omega_{{\rm 90}\%}<10\mathrm{deg}^2$, respectively.}
    \label{fig:sky_loc}
\end{figure*}

In Section \ref{sec:mm} we show the detection rates for joint GW and EM events considering representative detection limits based on the typical depth that can be reached during a wide-field EM follow-up in response to a GW alert in O4 and O5, without information on the GW sky localization. In order to provide the community with the opportunity to explore alternative observing configurations that correspond to varying detection limits as well as constraining the sky localization, we report in Fig. \ref{fig:rate_limit} and in Fig. \ref{fig:sky_loc}
the distribution of detection rates as a function of the detection limit for the joint EM+GW channels considered in this work, in O4 (upper panels) and O5 (lower panels). 

In particular, in Fig. \ref{fig:rate_limit} we show the detection rates for the GRB prompt+GW (left panel, orange) and GRB afterglow+GW (right panel, red) detectable binaries for our BHNS fiducial population (for KN\ae\, this information is already displayed in Fig. \ref{fig:kn}). For the GRB Prompt+GW detections, we showcase the rates under three different assumptions: we assume an all-sky field of view and a 100\% duty cycle (solid line), we account for the duty cycle and field of view of \textit{Fermi}/GBM (dashed line) and Swift/BAT (dotted line). Both in O4 and O5 the curves tend to flatten after a value of about 10 ph cm$^{-2}$ s$^{-1}$, as the GRB prompt detection is limited by the GW detection. For the GRB afterglow+GW we show individually the rates for the radio (solid), optical (dashed), and X (dotted) bands. The saturation point of the curves indicates the sensitivity required to detect all the EM+GW events, resulting in a corresponding detection rate for the GRB afterglow and KN (see Fig. \ref{fig:sky_loc}) of $0.34^{+0.34}_{-0.20}$ yr$^{-1}$ for the O4 run and $2.13^{+2.19}_{-1.25}$ yr$^{-1}$ for the O5 run (assuming the SFHo EoS and $\chi_\mathrm{BH} = 0$).

In order to provide additional information about the sky localization, in Fig. \ref{fig:sky_loc} we show the detection rates as function of the detection limit for the KN+GW (in the $g$ band) and GRB afterglow+GW (in the radio band), considering all the detectable binaries (solid lines), the detectable binaries with $\Delta\Omega_{{\rm 90}\%}<100$ $\mathrm{deg}^2$ (dashed lines) and the ones with $\Delta\Omega_{{\rm 90}\%}<10$ $\mathrm{deg}^2$ (dotted lines). In the top panel we assume the LVK O4 detectors network (colored lines) and a network consisting of only two aLIGO (black lines), while in the lower panel the LVKI O5 detectors network (colored lines) and the same network without LIGO–India (black lines).

In O4, as expected, there is a significant degradation in sky localization assuming a GW network without Virgo and KAGRA. However, the detection rates are still low for a complete network, due to the intrinsic challenges that we discussed previously, posed by the BHNS population. Regarding the O5 run, assuming the LVKI network, the sky localization is excellent, with nearly all events having $\Delta\Omega_{{\rm 90}\%}<100$ $\mathrm{deg}^2$. Furthermore, in the absence of LIGO-India, the presence of Virgo and KAGRA at their design sensitivities would result in only a twofold reduction in the rate relative to 100 square degrees, yielding a KN+GW rate of $\sim 1$ yr$^1$.

\end{appendix}
\end{document}